\documentclass[aps,pre,twocolumn,superscriptaddress,floatfix,notitlepage]{revtex4-2}
\usepackage[utf8]{inputenc}
\usepackage[english]{babel}
\usepackage{amsmath,amsfonts,amssymb,amsthm,bm,color,dsfont,graphicx,psfrag,mathtools,cancel}
\usepackage{hyperref}
\usepackage{tikz}
\usetikzlibrary{positioning}
\usepackage{pgfplots}
\pgfplotsset{compat=1.17}
\usepackage{siunitx}
\usepackage{xcolor}
\usepackage{soul}
\usepackage{braket}

\usepackage[export]{adjustbox}
\usepackage{natbib}
\usepackage[hybrid]{markdown}
\usepackage{mathrsfs}

\usepackage{blkarray}
\usepackage{multirow}
\usepackage[caption=false]{subfig}
\usepackage[normalem]{ulem}

\begin{document}

\title{{Signatures of the Quantum Geometric Dipole of Interlayer Excitons in Counterflow
Conductivity}}

\author{Fanuel I. Mendez}
\affiliation{Department of Physics, Indiana University, Bloomington, Indiana 47405, USA}
\affiliation{Quantum Science and Engineering Center, Indiana University, Bloomington, Indiana 47405, USA}

\author{Luis Brey}
\affiliation{Instituto de Ciencia de Materiales de Madrid (CSIC), Cantoblanco, 28049 Madrid, Spain}

\author{H.A. Fertig}
\affiliation{Department of Physics, Indiana University, Bloomington, Indiana 47405, USA}
\affiliation{Quantum Science and Engineering Center, Indiana University, Bloomington, Indiana 47405, USA}

\date{\today}

\begin{abstract}
{Collective excitations of many-body electron systems can carry internal structure, supporting novel quantum geometric and topological properties. Among these are  a quantum geometric dipole (QGD), which for excitons have direct significance as an internal polarization.  For interlayer excitons of a bilayer system, this represents an in-plane dipole moment, which can be used to drive them with in-plane electric fields. In this work, we consider  counterflow electric currents associated with driven excitons in such a bilayer system as a probe of their QGD structure.  As a simple but non-trivial example, we analyze a structure with a one-dimensional periodic potential in a strong perpendicular magnetic field. The resulting magnetoexciton bands host QGD structure that distinguishes it from the exciton QGD of a uniform system.  To model exciton transport
we adopt a Boltzmann approach that includes inter-band tunneling, allowing us to consider
non-equilibrium momentum distributions that result from strong layer-antisymmetric driving fields.  We show how linear response to a layer-symmetric component of the driving fields provide information about the QGD, and that the broad QGD structure of the exciton bands can be probed by the varying the layer-antisymmetric field.  Our results demonstrate that counterflow conductivity serves as a tunable probe of the internal quantum geometric structure carried by the interlayer excitons, connecting transport to the quantum geometry of many-body excitations.
}

\end{abstract}
\maketitle

\section{Introduction}
\label{Section1}

The emergence of two-dimensional Van der Waals (vdW) materials has in recent years allowed the realization of a number of exotic electron states in a variety of  multilayer systems \cite{geim_van_2013,novoselov_2d_2016, novoselov2005two, liu_van_2016}. These systems host striking behaviors made possible by the effects of electron-electron interactions on the system, including correlated insulators, superconducting states, magnetic states, and integer and fractional Chern insulators \cite{cao_correlated_2018, cao_unconventional_2018, huang_layer_2017, gong_discovery_2017, cai_signatures_2023, park2023observation, zeng_thermodynamic_2023, kennes_moire_2021, gibertini_magnetic_2019}.  In many of these states, quantum geometry plays a crucial role, and the unusual behavior this brings is enriched by interactions among the electrons.   The role of these interactions can be important both within and across layers.

For insulating states  -- typically semiconductors -- a fundamental marker of interaction effects is the presence of exciton states in the low-energy spectrum of the system  \cite{rohlfing_electron-hole_2000, wang_colloquium_2018}. Excitons are bound particle-hole excitations akin to hydrogen atoms, in which the electronic structure of the electron and hole bands have important qualitative effects \cite{wannier_structure_1937, knox1963theory}. In two-dimensional materials such as van der Waals layers, they are often particularly low in energy (i.e., more strongly bound) than is the case for bulk materials, due to the relatively weak screening of the Coulomb interaction by the material itself \cite{chernikov2014exciton, mueller_exciton_2018}. While band energetics have long been known to impact exciton energies and behaviors, it has more recently been appreciated that the quantum geometries of these bands impact the exciton properties as well \cite{srivastava_signatures_2015, zhou_berry_2015}.  Indeed, as collective modes with well-defined momenta, exciton states have their own quantum geometries \cite{yao_berry_2008, kwan_exciton_2021}, and because of their internal structure they admit a richer set of quantum geometric characterizations than is possible for single-electron bands \cite{cao_quantum_2021}.

In this work, we focus on a unique quantum geometric property of excitons: their quantum geometric dipole (QGD) \cite{fertig_many-body_2025}.  This is an internal polarization of the exciton that is tied to its momentum in the Hilbert space of an exciton band \cite{chaudhary2021anomalous, paiva2024shift, davenport2026exciton, xie_theory_2024, lozano_optical_2025}.  Direct analogs of this quantum geometric structure may also be found in magnons \cite{chen_quantum-geometric_2025} and plasmons \cite{cao2021quantum, cao2022plasmonic}.  Of particular interest in this context are \textit{interlayer} excitons, in which the electron and hole reside in different layers of a multilayer structure. The segregation of the constituents into these separate spaces endows the exciton with attractive properties and possibilities: control of its energy with perpendicular electric field \cite{jauregui_electrical_2019}, enhanced lifetime due to the physical separation between the particle and hole \cite{rivera_observation_2015}, and the possibility of applying different electric fields in the two layers, yielding different forces on the electron and hole \cite{eisenstein1990independently,eisenstein_macdonald_2004, liu_quantum_2017, li_excitonic_2017}. In vdW materials, such interlayer excitons have been realized \cite{jauregui_electrical_2019, rivera_observation_2015, wang_evidence_2019} in pairs of transition metal dichalcogenide (TMD) layers separated by one or more layers of hexagonal boron nitride (hBN), for which tunneling between the layers is negligibly small. Analogous systems may be fabricated using graphene layers to host electrons and holes, with a strong magnetic field introducing gaps in their electronic spectra \cite{liu_quantum_2017, li_excitonic_2017}.  In this latter case, the relevant excitations are magnetoexcitons.



\begin{figure*}[t]
    \centering
    \subfloat[]{\includegraphics[width=0.48\textwidth, height=0.6\textheight, keepaspectratio]{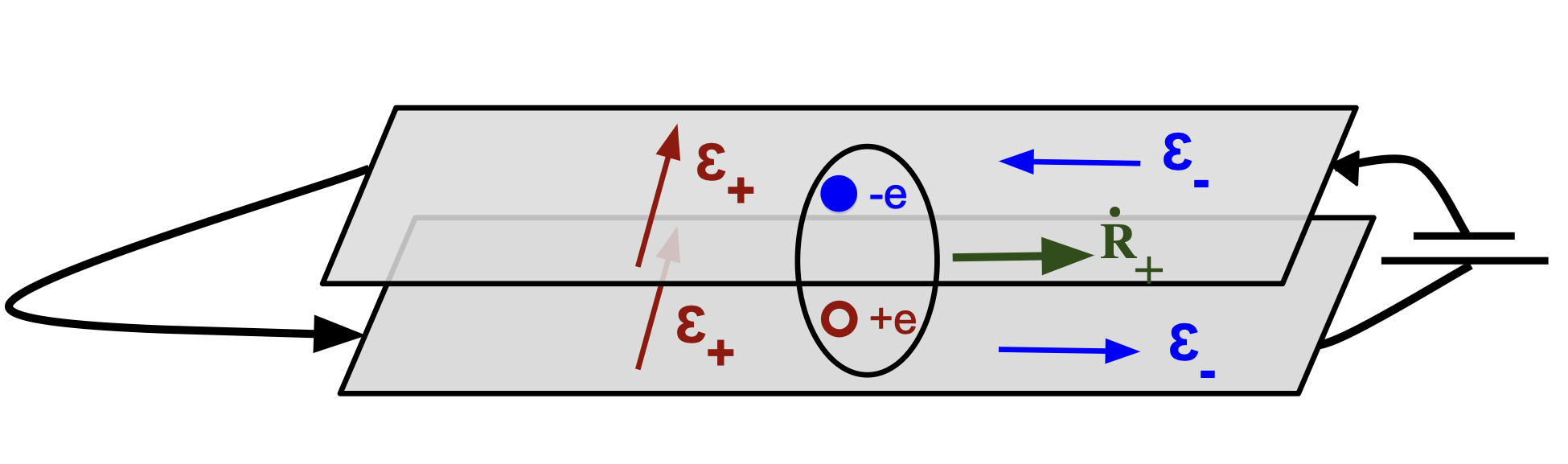}
    \label{fig:schematic_a}}
    \hspace{0.02\textwidth}
    \subfloat[]{\includegraphics[width=0.48\textwidth, height=0.15\textheight, keepaspectratio]{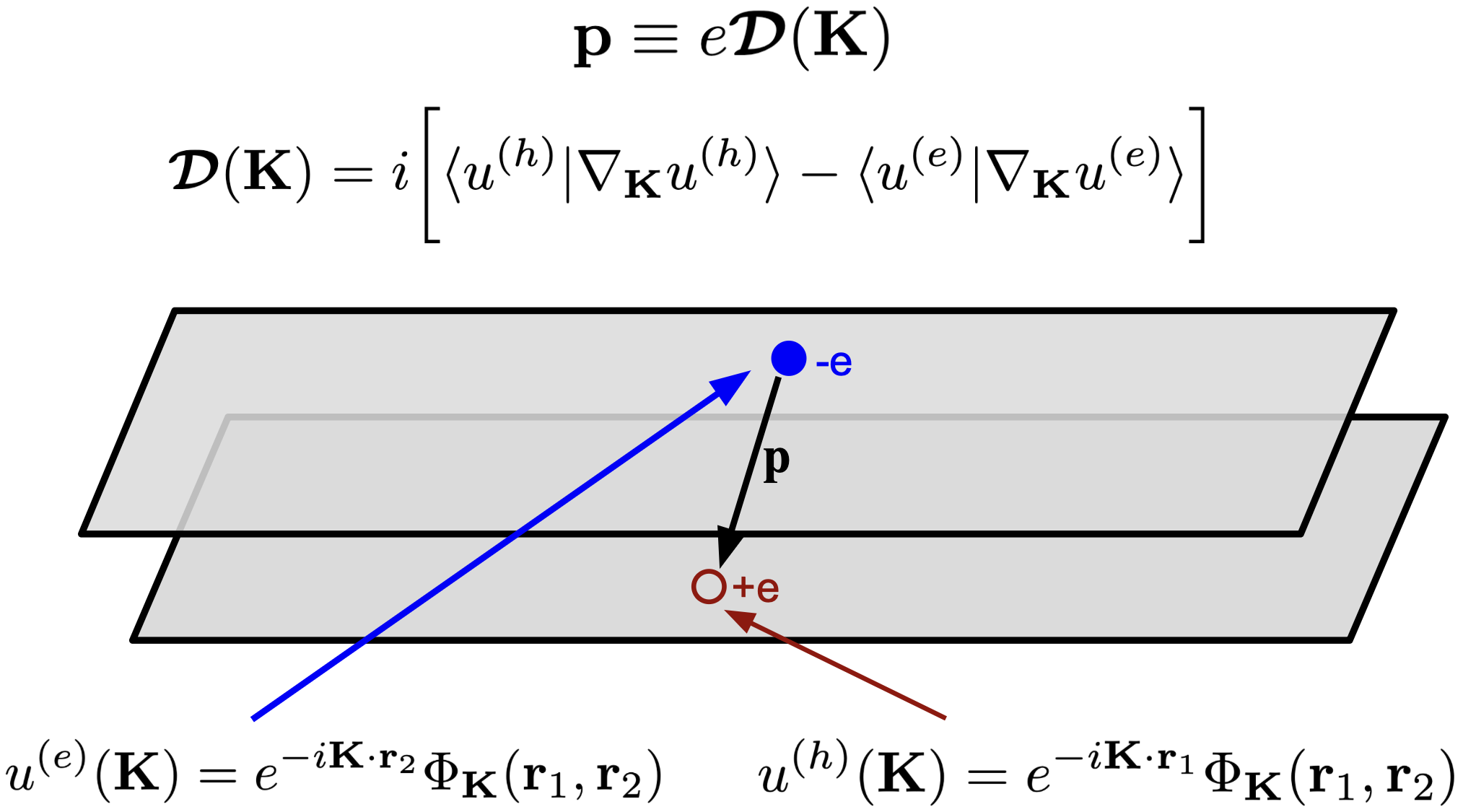}
    \label{fig:schematic_b}}
    \caption{(a) Schematic of interlayer exciton counterflow current in a bilayer system. Electric fields applied individually to the electron and hole layers give rise to  symmetric ($\boldsymbol{\mathcal{E}}_{+}$) and antisymmetric ($\boldsymbol{\mathcal{E}}_{-}$) fields. In semiclassical equations of motion, the symmetric field couples to the quantum geometric dipole and drives the center of mass velocity $\dot{{\bf R}}_{+}$ whereas $\boldsymbol{\mathcal{E}}_{-}$ induces time evolution of the exciton center-of-mass momentum, impacting the exciton dipole moment. (b) Schematic of the dipole moment in a bilayer system. The  quantum geometric dipole moment is a gauge-invariant quantity given by the difference of the Berry connections of the hole and electron, constructed from cell-periodic exciton states $u^{(h,e)}({\bf{K}}) = e^{-i{\bf K}\cdot{\bf r}_{1,2}}\Phi_{{\bf K}}({\bf r}_1,{\bf r}_2)$, where $\Phi_{{\bf K}}({\bf r}_1, {\bf r}_2)$ is the exciton wave function and ${\bf r}_1$ (${\bf r}_2$) are the spatial positions of the hole (electron).}
    \label{fig:schematic}
\end{figure*}

\begin{figure*}[t]
    \centering
    \begin{minipage}[t]{0.45\textwidth}
        \centering
        \subfloat[]{\includegraphics[width=\textwidth, height=0.6\textheight,keepaspectratio]{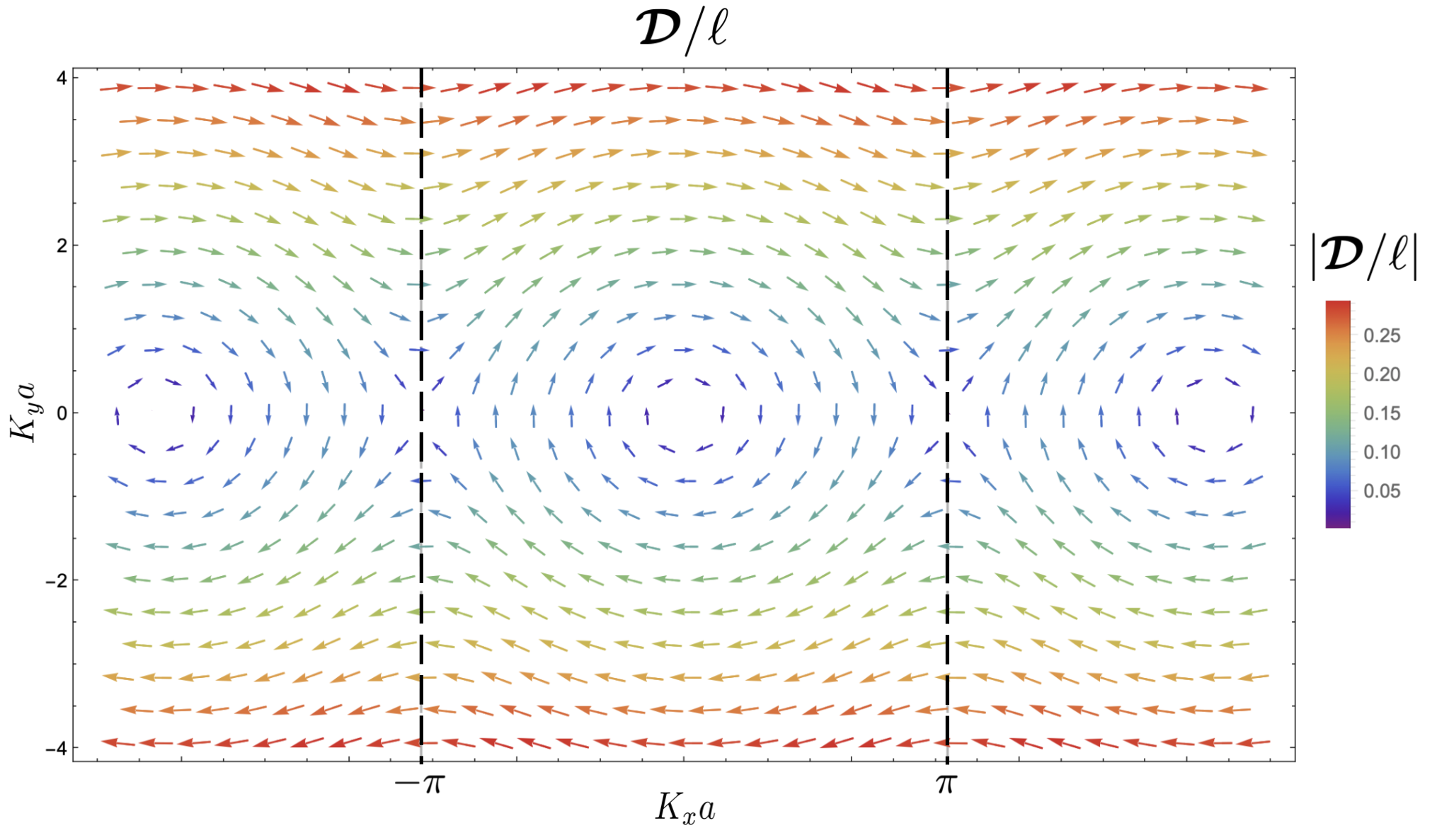}
        \label{fig:QGD_uniform}}
    \end{minipage}
    \hspace{0.02\textwidth}
    \begin{minipage}[t]{0.45\textwidth}
        \centering
        \subfloat[]{\includegraphics[width=\textwidth, height=0.2\textheight, keepaspectratio]{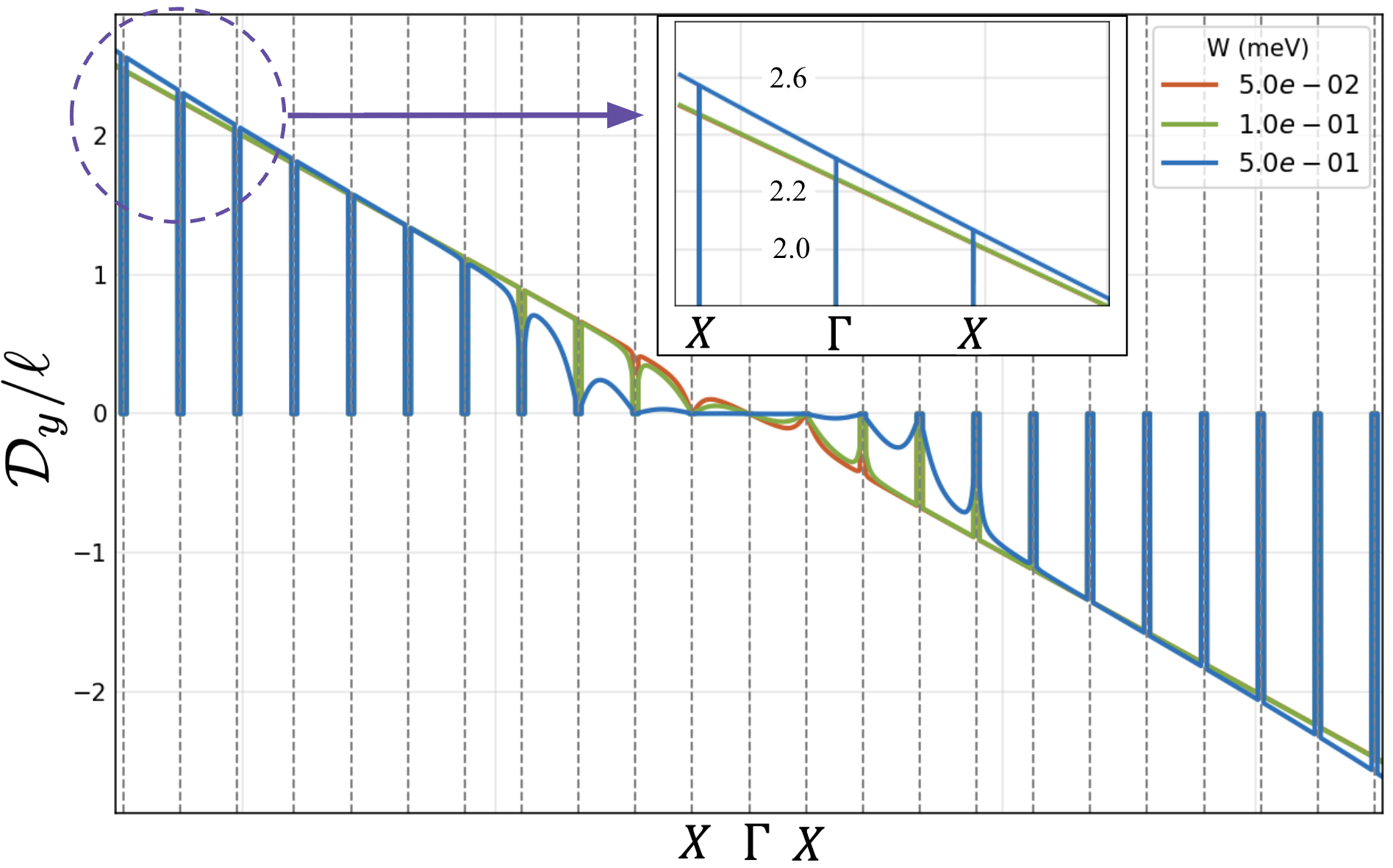}
        \label{fig:QGD_W}}
    \end{minipage}
    \caption{Quantum geometric dipole $\boldsymbol{\mathcal{D}}_{n}({\bf {K}})$ of a magnetoexciton in a strong magnetic field in a periodic potential. (a) Vector plot of the QGD in a bilayer system in a unidirectional periodic potential for the lowest band ($n=0$). (b) Band-projected QGD for fixed $K_y = 0$ for varying periodic potential strengths $W$ (meV) in the extended zone scheme. In (b), the dashed vertical lines indicate avoided crossings where the QGD vanishes. Inset: Behavior of the QGD for large momentum magnitude in extended zone scheme. Parameters for (a): $\ell = 100 \,\si{\angstrom}$, $d = 3 \, \text{nm}$, $a = 140 \, \text{nm}$ , $\kappa = 3.76$, $W = 0.1$ meV and $N_c = 2$. Parameters for (b): $\ell = 100 \,\si{\angstrom}$, $d = 3 \, \text{nm}$, $a = 140 \, \text{nm}$ , $\kappa = 3.76$  and $N_c = 20$. }
    \label{fig:QGD}
\end{figure*}

The connection between the electric dipole moment of a magnetoexciton in the strong field limit and its momentum is well-known\cite{kallin_excitations_1984, lerner_mott_1980, khalatnikov_two-dimensional_1996, kozin2021anomalous}, but more recently this connection has been reinterpreted and generalized to other systems from a quantum geometric perspective \cite{cao_quantum_2021, yang2026giant}.  One consequence of this is that the semiclassical dynamical equation of motion of an exciton wave packet is influenced by quantum geometry not just through the anomalous velocity associated with the Berry's curvature, but also through a drift velocity \cite{cao_quantum_2021, chaudhary2021anomalous} associated with the QGD.  The consequences of this are largely unexplored and are the subject of this study.  Specifically, we  develop a Boltzmann approach to counterflow conductivity carried by a collection of interlayer excitons, focusing on contributions from the QGD.  By exploiting the freedom to consider different in-plane components of the electric field in each layer (see Fig. \ref{fig:schematic}), we show it is possible to vary the momentum distribution of the interlayer excitons (through a  strong layer-antisymmetric component of the electric field) while isolating the contribution of the QGD to counterflow transport (via linear response to a layer-symmetric electric field component).  We solve these equations for the specific case of a collection of interlayer magnetoexcitons in the strong magnetic field limit, subject to a unidirectional periodic potential that introduces an exciton band structure with an associated QGD profile for each band (see Fig. \ref{fig:QGD}.)  We show that the counterflow conductivity provides information on how the QGD evolves as higher bands are occupied, through increasingly strong antisymmetric driving fields. We note that, analogously, semiclassical transport has been explored as an indicator for Berry's curvature effects in both Bloch electrons and excitons \cite{xiao2010berry, nagaosa2010anomalous, xiao2005berry, sodemann2015quantum,guo2025linear,jauregui2017curved,kovalev2019quantum}.  Indeed, non-linear transport of electrons through a band structure has recently been proposed as a probe that indirectly maps its Berry's curvature profile \cite{de_beule_berry_2023}.



Transport of interlayer excitons naturally leads to counterflow currents in a bilayer: when moving in real space, the motion of the particle and hole constituents in their respective layers represents oppositely directed electric currents 
\cite{finck_exciton_2011, su_how_2008, kellogg2002observation, kellogg2004vanishing, tutuc2004counterflow,wiersma2004activated}.  The transverse component of this counterflow current is quantified by $\sigma_{xy}^{(CF)}$, the linear response to an interlayer symmetric electric field.  As we show below, this quantity is an average of the {\it derivative} of the QGD with respect to the crystal momentum $K_x$ along the periodic potential axis, and it depends non-trivially on both the strength of the periodic potential and the antisymmetric component of the electric field.    In general, we find that increasing the periodic potential magnitude tends to suppress $\sigma_{xy}^{(CF)}$, in accord with the accompanying suppression of the QGD one finds for the low-lying bands (see Fig. \ref{fig:QGD_W}.)  By contrast, increasing the layer - antisymmetric driving field enhances $\sigma_{xy}^{(CF)}$, as the exciton population moves to higher energy states, which on average host higher QGD slopes.  In this way, the transverse counterflow conductivity provides information about the QGD structure hosted by the exciton bands.

The remainder of this article is organized as follows. In Sec.~\ref{subQGD} we review the QGD and its connection to the exciton electric dipole moment.  Section \ref{SemiclassicalEq} presents the semiclassical equations of motion for the band-projected exciton dynamics, and how the QGD of interlayer magnetoexcitons of a two-dimensional bilayer in the presence of a unidirectional periodic potential enters them.  In Section \ref{CFConduct} we discuss counterflow conductivity from exciton transport and how this can be described in terms of the exciton QGD and a Boltzmann distribution of excitons. We complete our formal discussion in Section \ref{BoltzDis}, where we explain our approach to computing the multi-band Boltzmann distribution, including interband tunneling in the model.  
Section \ref{ResultsandDisc} presents numerical results that emerge from our formalism.  We conclude with a summary and discussion of our results in Section \ref{SecConclusion}.
In addition, we provide details of our calculations in two appendices.  Appendix \ref{AppendixBandStructure} provides details of our band structure and QGD calculations. Appendix \ref{AppendixBoltzmann} supplies further details of our counterflow conductivity calculations.

\section{General Formulation}
\label{Section2}
\subsection{Quantum Geometric Dipole}
\label{subQGD}
We begin by briefly reviewing the derivation of the QGD and its connection to the internal electric dipole moment of an exciton.  We define $\Phi_{{\bf K}}({\bf r}_1, {\bf r}_2) = \braket{{\bf r}_1,{\bf r}_2|\Phi_{{\bf K}}}$ as the two-body wavefunction for the exciton, where ${\bf r}_1$ and ${\bf r}_2$ are the  hole and electron positions and ${\bf K}$ is the total exciton momentum. Cell-periodic functions may be constructed via a family of mappings of the form
\begin{equation}
    \ket{u_{{\bf K}},\alpha} \equiv e^{-i[\alpha {\bf r}_1 + (1-\alpha){\bf r}_2] \cdot {\bf K}}\ket{\Phi_{\bf K}}.
\end{equation}
This construction 
takes account of the internal structure of an exciton state, with specific form determined by the parameter $\alpha$.  For $\alpha = 1 (0)$ the connection associates the Bloch phase solely with the hole (electron),
allowing one to define Berry's connections specific to each constituent,
\begin{equation}\label{Berryconnhole}
    \boldsymbol{\mathcal{A}}^{(h)}({\bf K}) \equiv i\braket{u_{{\bf K}},1|\nabla_{{\bf K}}|u_{{\bf K}},1},
\end{equation}
\begin{equation}\label{Berryconnele}
    \boldsymbol{\mathcal{A}}^{(e)}({\bf K}) \equiv i\braket{u_{{\bf K}},0|\nabla_{{\bf K}}|u_{{\bf K}},0},
\end{equation}
where $\boldsymbol{\mathcal{A}}^{(e)}({\bf K})$ and $\boldsymbol{\mathcal{A}}^{(h)}({\bf K})$ are the connections associated with electron and hole, respectively. The electric dipole moment may then be shown \cite{cao_quantum_2021} to obey ${\bf{p}} \equiv e \boldsymbol{\mathcal{D}}({\bf K})$, where 
$\boldsymbol{\mathcal{D}} ({\bf K})\equiv \boldsymbol{\mathcal{A}}^{(h)}({\bf{K}}) - \boldsymbol{\mathcal{A}}^{(e)}({\bf{K}})$.  Because of its quantum geometric construction, the latter quantity is referred to as the quantum geometric dipole (QGD).  
Note that, while (as usual) the Berry connections are gauge-dependent, their difference is gauge-invariant. Thus, the in-plane dipole moment of a two-dimensional exciton may be understood as being entirely determined by the quantum geometry of the excitonic wavefunctions, as illustrated in Fig. \ref{fig:schematic_b}. The QGD produces a correction to the exciton velocity beyond that of the Berry's curvature, as we discuss below.

While the existence of an internal dipole moment is well-established for magnetoexcitons in a strong magnetic field \cite{cao_quantum_2021,fertig_many-body_2025}, its interpretation in terms of quantum geometric quantities suggests that it is not limited to this setting. We emphasize that the formulation of $\boldsymbol{\mathcal{D}}({\bf K})$ is quite general, and applies to any collective mode that can be labeled by and continuously evolves with a well-defined momentum ${\bf K}$, making it a broadly applicable quantum geometric quantity \cite{fertig_many-body_2025}. Nevertheless, because of its relative analytical simplicity, in what follows we focus on a system of interlayer magnetoexcitons in a two-layer system, subject to a unidirectional periodic potential. Exciton currents in such systems naturally manifest counterflow electric currents \cite{finck_exciton_2011, kellogg2002observation, tutuc2004counterflow}, and our goal in this study is to understand how exciton QGD's impact these currents.  We thus turn next to their dynamics.

\subsection{Semiclassical Equations of motion}
\label{SemiclassicalEq}

Our starting point is the band-projected semiclassical theory of interlayer exciton dynamics in a two-layer system. The equations of motion for the center of mass position ${\bf{R}}_+$ and average momentum ${\bf{K}}$ of an exciton wavepacket are given by \cite{cao_quantum_2021}
\begin{equation}\label{EOMposition}
    \hbar\dot{\bf{R}}_{+,n} + \hbar\dot{\bf{K}}\times {\bf{\Omega}}_{n}({\bf{K}}) = 2\nabla_{\bf{K}}E_{n}({\bf{K}}) +e\nabla_{{\bf{K}}}[\boldsymbol{\mathcal{E}}_{+}\cdot \boldsymbol{\mathcal{D}}_{n}({\bf{K}})],
\end{equation}
\begin{equation}\label{EOMmomentum}
    \hbar \dot{{\bf{K}}} = -e\boldsymbol{\mathcal{E}}_{-},
\end{equation}
where $\boldsymbol{\Omega}_{n}({\bf K}) = \vec{\nabla}\times \left(\boldsymbol{\mathcal{A}}_{n}^{(h)}({\bf{K}}) + \boldsymbol{\mathcal{A}}_{n}^{(e)}({\bf{K}})\right)$ 
is the effective Berry's curvature for the exciton, $\boldsymbol{\mathcal{E}}_{\pm} = \boldsymbol{\mathcal{E}}_{h} \pm \boldsymbol{\mathcal{E}}_{e}$ are the symmetric and antisymmetric combinations of the electric fields applied to the hole and electron layers respectively, $E_{n}({\bf{K}})$ is the exciton energy dispersion, and $\boldsymbol{\mathcal{D}}_{n}({\bf{K}})$ is the exciton quantum geometric dipole.  Note in these expressions, the subscript $n$ is a band index. It is important to note that the exciton equations of motion involve three quantities that are specific to the particular system they are describing: the energy dispersion $E_{n}({\bf{K}})$, the Berry's curvature ${\bf{\Omega}}_{n}({\bf{K}})$, and the quantum geometric dipole $\boldsymbol{\mathcal{D}}_{n}({\bf{K}})$. In the case of magnetoexcitons in a strong magnetic field, for exciton constituents confined to the lowest Landau levels, a non-trivial band structure can be induced via the insertion of a periodic potential, which we assume to be unidirectional.  As a simple model, we adopt for this potential the form
\begin{equation}\label{potential}
     V({\bf{r}}_1,{\bf{r}}_2) = W\sum_{i=1}^{2}\cos({\bf{g}}\cdot{\bf{r}}_{i}),
\end{equation}
where ${\bf{g}} = 2\pi\hat{x}/a$, $a$ is the superlattice constant, $W$ is the strength of the modulation potential, and ${\bf{r}}_1$ (${\bf{r}}_2$) are the two-dimensional hole (electron) positions in each layer.

Among the quantities needed for the semiclassical equations of motion are the QGD's for the different bands, which may be shown to have the form (see Appendix \ref{AppendixBandStructure})
\begin{equation}\label{MultiQGD}
   \boldsymbol{\mathcal{D}}_{n}({\bf{K}}) = \sum_{m = -N_c}^{N_c}|c_{m}^{(n)}({\bf{K}})|^{2}\boldsymbol{\mathcal{D}}_{ME}({\bf{K}} +m{\bf{g}}),
\end{equation}
where $\boldsymbol{\mathcal{D}}_{ME}({\bf q}) \equiv {\bf q} \times\hat{z} \ell^2$ is the  QGD of a magnetoexciton for a uniform two-dimensional electron gas \cite{cao_quantum_2021,fertig_many-body_2025}, $N_c$ is a cutoff chosen so that the results are not noticeably affected if made larger, and $c_{m}^{(n)}({\bf{K}})$ are expansion coefficients of the \textit{n}$^{th}$ exciton band eigenstate in terms of states in the absence of the periodic potential. 

Representative QGD's for various strengths of the periodic potential are illustrated in Fig. \ref{fig:QGD}. In the full two-dimensional momentum space, one observes for the QGD of the lowest band in a repeated zone scheme (Fig. \ref{fig:QGD_uniform}) a helical structure \cite{yang2026giant} near the origin which is inherited from the form of $\boldsymbol{\mathcal{D}}_{ME}$.  In addition, the QGD at a Brillouin zone (BZ) boundary is perpendicular to it, a requirement of the identification of momentum points separated by the reciprocal lattice vector ${\bf g}$.  A broader view of the QGD structure may be seen in an extended zone scheme, as illustrated in Fig. \ref{fig:QGD_W} for $K_y=0$.  Three features are noteworthy for this value of $K_y$. ({\it i}) The QGD vanishes at a BZ boundary.  This occurs due to the periodicity of the wavefunctions across it.  Its complete elimination is specific to $K_y=0$; more generally the component of $\boldsymbol{\mathcal{D}}_{n}({\bf{K}})$ parallel to a BZ boundary vanishes.  ({\it ii}) For stronger periodic potentials (larger $W$ in our model), the QGD tends to be suppressed for the lowest bands.  This is an indication that for the lowest energy states, the exciton wave functions vary only weakly with {\bf K}, keeping a form that tends to minimize the potential energy term $V({\bf r}_1,{\bf r}_2)$.  Note that the energy dispersion is relatively flat for such states as well.  (The computation of exciton band structure energies is described in Appendix  \ref{AppendixBandStructure}.) ({\it iii}) For high bands, away from BZ boundaries, the QGD tends to develop a larger magnitude than would be expected based purely on zone-folding of $\boldsymbol{\mathcal{D}}_{ME}$.  This occurs because the energy dispersion of exciton states at high ${\bf K}$ tends to weaken as the electron and hole become increasingly remote from one another.  A coarse-grained density of states thus increases with energy, so that there are effectively more states of high total momentum than of low total momentum admixed into a state at some crystal momentum {\bf K}.  Thus, the distribution of  $\boldsymbol{\mathcal{D}}_{ME}({\bf K} +{m\bf g} )$ values contributing to $\boldsymbol{\mathcal{D}}_{n}({\bf K})$ in Eq. \ref{MultiQGD} skew  towards large magnitudes as the band index increases.  Because of the underlying mechanism, the effect becomes more pronounced as the periodic potential increases relative to the electron-hole interaction.  This is evident in the inset to Fig. \ref{fig:QGD_W}.



We next turn to our analysis of the counterflow conductivity from a collection of excitons, and explain how these features impact its behavior.

\subsection{Counterflow Conductivity}
\label{CFConduct}
Our approach to the computation of the counterflow conductivity proceeds using a Boltzmann equation approach.  In practice, this means we focus on situations in which
the excitons are sufficiently dilute and the temperature is sufficiently high that interactions among them may be ignored.  Note this means we do not include the possibility of Bose-Einstein condensation of the excitons in our analysis.  We consider a simple geometry in which there is a uniform spatial distribution of excitons, but a drift current flows due to an externally applied layer-symmetric electric field.  This exciton current is controlled by a Boltzmann distribution function $f_n({\bf K})$ which describes the density of excitons in ${\bf K}$-space for the $n^{th}$ band, as well as the equation of motion Eq. \ref{EOMposition}.  

The counterflow (electric) current associated with real-space motion of the excitons may be expressed as 
\begin{equation}\label{excitoncurrent}
    {\bf{j}}_{+}^{(CF)} = e\sum_{n}\int_{BZ} \frac{d^2K}{(2\pi)^2}f_{n}({\bf{K}})\dot{\bf{R}}_{+,n}.
\end{equation}
Note that the distribution is a function of the layer antisymmetric electric field $\boldsymbol{\mathcal{E}}_{-}$.  This parameter plays an important role in what follows, in that we use it to manipulate the exciton distribution, allowing the excitons to explore QGD effects from different regions of the band structure.   Since we are solely interested in the QGD effects on exciton transport, we define $\delta {\bf{j}}_{+}^{(CF)}(\boldsymbol{\mathcal{E}}_{+}, \boldsymbol{\mathcal{E}}_{-}) \equiv {\bf{j}}_{+}^{(CF)}(\boldsymbol{\mathcal{E}}_{+}, \boldsymbol{\mathcal{E}}_{-}) - {\bf{j}}_{+}^{(CF)}(0, \boldsymbol{\mathcal{E}}_{-})$;  one sees from Eqs. \ref{EOMposition} and \ref{EOMmomentum} that this quantity encodes the contribution of the QGD to the motion of an exciton.  The counterflow current that emerges from this quantity is a linear response to the symmetric field $\boldsymbol{\mathcal{E}}_{+}$ in the presence of the static antisymmetric electric field $\boldsymbol{\mathcal{E}}_{-}$. Defining a counterflow conductivity matrix via the relation $\delta {\bf{j}}_{+}^{(CF)} = \overleftrightarrow{\sigma}^{(CF)}\boldsymbol{\mathcal{E}}_{+}$, its components can be expressed as (see Appendix \ref{AppendixBoltzmann}) 
\begin{equation}\label{SigmaFinalExp}
    \sigma_{\nu\mu}^{(CF)} = \frac{e^2}{\hbar}
    \sum_{n}\int_{BZ}\frac{d^2K}{(2\pi)^2}
    f_{n}({\bf{K}})\frac{\partial \mathcal{D}_{n,\mu}}
    {\partial K_{\nu}}, \quad \nu,\mu = x,y.
\end{equation}
As shown in Appendix B,
\begin{equation}\label{QGDxBand}
    \mathcal{D}_{n,x} = K_y \ell^2
\end{equation}
so that
\begin{equation}\label{SigmaYX}
    \sigma_{yx}^{(CF)}\equiv \frac{e^2\ell^2}{\hbar}n_{exc},
\end{equation}
where $n_{exc}$ is the exciton number density.  Interestingly, the difference in the magnitudes of $\sigma^{(CF)}_{xy}$ and $\sigma^{(CF)}_{yx}$ directly reflects the impact of the periodic potential on the QGD structure of the excitons. Finally, we note that the diagonal components of the counterflow conductivity are strongly suppressed: $\sigma_{xx}^{(CF)}$ vanishes as a direct consequence of Eq.~\eqref{QGDxBand}, while $\sigma_{yy}^{(CF)}$ is nonzero but negligibly small compared to $\sigma^{(CF)}_{xy}$ and $\sigma^{(CF)}_{yx}$ throughout the parameter space considered.

\begin{figure}[t]
    \centering
    \includegraphics[width=0.46\textwidth]{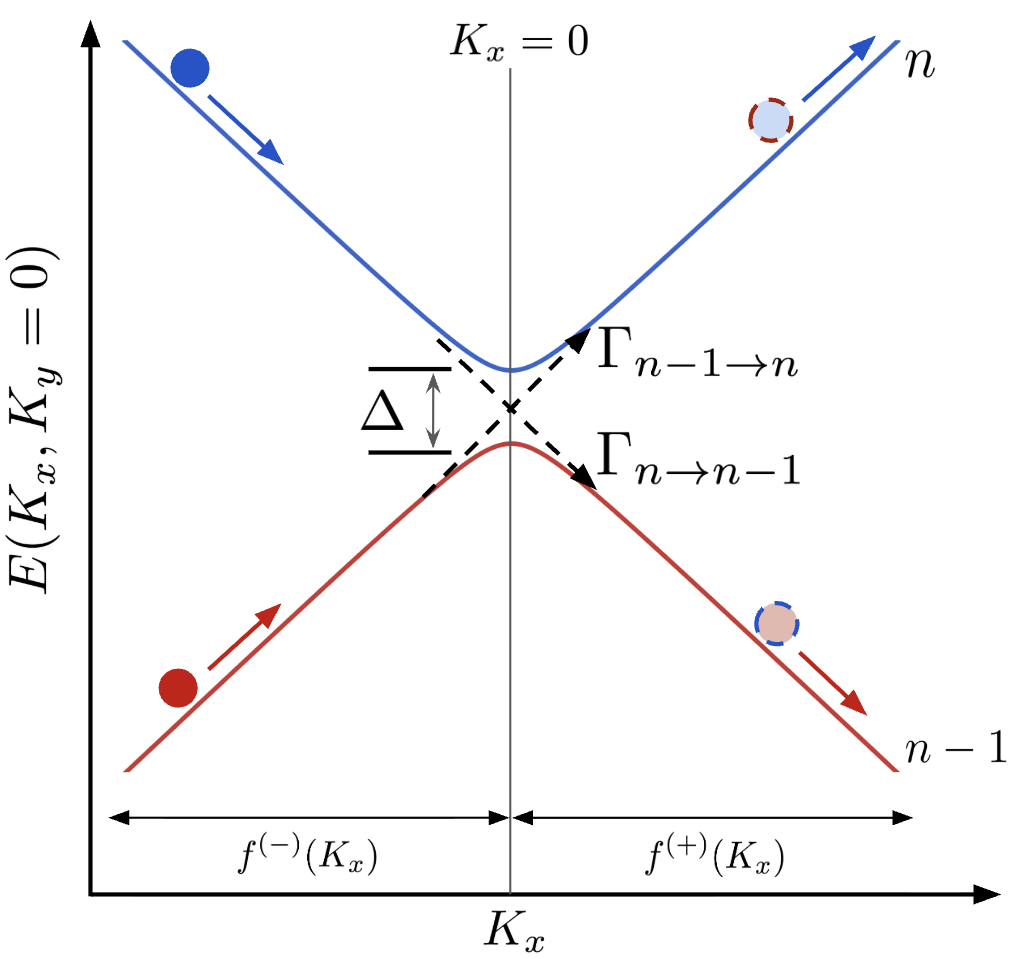}
    \caption{Schematic an avoided crossing between two exciton bands at $K_x = 0$ with gap $\Delta$. Arrows along each band indicate the direction of the exciton flow driven by $\mathcal{E}_{-}$. Dashed arrows crossing the gap represent the interband tunneling rates $\Gamma_{m \rightarrow n}$ and $\Gamma_{n \rightarrow m}$. The distribution is divided into regions $f^{(-)}(K_x)$ and $f^{(+)}(K_x)$ corresponding to $K_x <0$ and $K_x > 0$, respectively.}
    \label{fig:LZTunnel}
\end{figure}
\subsection{Boltzmann Distribution}
\label{BoltzDis}
The quantity of interest is thus $\sigma_{xy}^{(CF)}$, and from Eq. \ref{SigmaFinalExp} it is clear one needs a concrete form for the nonequilibrium  distribution function $f_n({\bf k})$ to compute it.  We do so using a steady-state time-independent Boltzmann transport equation in the relaxation-time approximation, adapted to a multi-band setting.  Crucially, we allow excitons to tunnel between bands at avoided crossings in the band structure, capturing the very different behaviors one expects for rapid or slow transit of excitons across regions of near degeneracy in the band structure.  With input from Eq. \ref{EOMmomentum}, our phenomenological Boltzmann equation takes the form

\begin{widetext}
\begin{multline}
    -\frac{e}{\hbar}\boldsymbol{\mathcal{E}}_{-}\cdot \nabla_{{\bf{K}}} f_{n}({\bf{K}}) = 
    \frac{f_{n}^{0}({\bf{K}}) - f_{n}({\bf{K}})}{\tau} 
    + \Gamma_{n-1\rightarrow n}({\bf{K}}) + \Gamma_{n+1\rightarrow n}({\bf{K}})
    - \Gamma_{n\rightarrow n-1}({\bf{K}}) - \Gamma_{n\rightarrow n+1}({\bf{K}}),
    \label{eq:boltzmann}
\end{multline}
where $\tau$ is the momentum-relaxation time and $f_{n}^{0}({\bf{K}})$
is the equilibrium distribution, which we take to have a simple
Maxwell-Boltzmann form.  Note that we have not introduced generation or recombination terms, in which case we are completely ignoring tunneling of of electrons and/or holes between the layers. To focus on the  physics of interest, we assume $\boldsymbol{\mathcal{E}}_{-}$ is applied along the superlattice axis (x-axis).  With this choice, Eq. \ref{eq:boltzmann} becomes a one-dimensional equation for each $K_y$.

The elements of this equation which are unusual are the
$\Gamma_{m \to n}({\bf K})$ terms, which represent tunneling rates between bands $m$ and $n$, assumed to occur at an avoided crossing located at $ {\bf K}$ \cite{yan2009observation}.  In the unidirectional periodic structure, avoided crossings occur only at ${\bf{K}} = (0,K_y)$ and ${\bf{K}} = (\pm\pi/a,K_y)$, which we refer to as the $0$ and $X$ points, respectively, and we exclusively consider tunneling between bands that actually anticross, so that non-zero values of $\Gamma_{m \to n}({\bf K})$ are retained only when $|m-n|=1$.  Our model for these tunneling rates is based on Landau-Zener tunneling, for which the probability of a system jumping between tunnel-coupled bands $m$ and $n$ is 
\cite{rubbmark1981dynamical}
\begin{equation}
    P_{mn}(K_x) = e^{-2\pi \frac{|\Delta_{mn}(K_x)|^2}{\hbar v(K_x)}},
\label{eq:LZ}
\end{equation}
where $\Delta_{mn}(K_x)$ is the band gap, and $v(K_x) = \frac{e}{\hbar}\frac{\partial E}{\partial K_x}\mathcal{E}_{-}$ is the  rate at which the exciton passes through the avoided crossing (see Eq. \ref{EOMmomentum}), which we refer to 
as the slew rate, for concreteness taken as positive in our analysis. Note that $\Delta_{mn}(K_x)$ is always evaluated at $K_x=0$ or $K_x=X$.  To form tunneling rates from Eq. \ref{eq:LZ}, we divide the distribution function between regions, defining $f_n^{(+)}(K_x) \equiv f_{n}(0< K_x< \frac{\pi}{a})$ and $f_n^{(-)}(K_x) \equiv f_{n}(-\frac{\pi}{a} < K_x < 0)$, as illustrated schematically in Fig. \ref{fig:LZTunnel}. The direction of exciton flow along each band is indicated, and the dashed lines crossing the gap represent the interband tunneling rates at the avoided crossing. We then write 
\begin{equation}\label{TRmnPaper}
    \Gamma_{m \rightarrow n} (K_x) = \dot{K}_x\biggl[P_{mn}(0) \,f_{m}^{(-)}(0)\,\delta[K_x] + P_{mn}(X)\,f_{m}^{(+)}(X)\delta[K_x-X]\biggr],
\end{equation}
\begin{equation}\label{TRnmPaper}
    \Gamma_{n \rightarrow m} (K_x) = \dot{K}_x\biggl[P_{mn}(0) \,f_{n}^{(-)}(0)\,\delta[K_x] + P_{mn}(X)\,f_{n}^{(+)}(X)\delta[K_x-X]\biggr].
\end{equation}
Since away from the avoided crossings the band occupations evolve independently, Eq. \eqref{eq:boltzmann} takes the same form for each band $n$. It is therefore convenient to collect the distribution functions for all bands into a column vector $\vec{f}^{(\pm)}(K_x)$, whereupon Eq. \eqref{eq:boltzmann} takes the compact matrix form
\begin{equation}\label{BoltzAwayPaper}
    \biggl[\dot{K}_x\frac{\partial}{\partial K_x} + \frac{1}{\tau}\biggr] \vec{f}^{(\pm)}(K_x) = \frac{1}{\tau}\vec{f}_{eq}(K_x),
\end{equation}
where $\vec{f}_{eq}(K_x)$ is a column vector containing the Maxwell-Boltzmann equilibrium distribution. Using an integrating factor, the solution in each region takes the form
\begin{equation}\label{SolutionPaper}
    \vec{f}^{(\pm)}(K_x) = e^{-\frac{K_x}{\dot{K}_x\tau}}\biggl[\vec{f}^{(\pm)}(0) + \frac{1}{\dot{K}_x\tau}\int_{0}^{K_x} dK' \,e^{\frac{K'}{\dot{K}_x\tau}} \vec{f}_{eq}(K')\biggr].
\end{equation}
A key feature of this approximation scheme is that ${f}_n(K_x)$ is \textit{discontinuous} across the tunneling gaps when the excitons are driven by $\boldsymbol{\mathcal{E}}_{-}$.  This is encoded by the differences in $f_n^{(+)}(K_x)$ and $f_n^{(-)}(K_x)$ as $K_x$ approaches either $0$ or the $X$ point.  Specifically, integrating Eq. \eqref{eq:boltzmann} around $K_x = 0$ yields the matching condition
\begin{equation}\label{M0equationPaper}
    \vec{f}^{(+)}(0) = M_{0}\,\vec{f}^{(-)}(0),
\end{equation}
and integrating around $K_x=X$ yields
\begin{equation}\label{MXequationPaper}
    \vec{f}^{(-)}(X) = M_{X}\,\vec{f}^{(+)}(X),
\end{equation}
where $M_0$ and $M_x$ are $N \times N$ tridiagonal matrices and $N = 2N_c +1$ is the total number of bands retained in the calculation. Convergence is assessed by monitoring the tails of the Boltzmann distribution function- when increasing $N$ leaves the tails essentially unaffected, the truncation is considered sufficient and the higher bands are found not to contribute noticeably to the counterflow conductivity.

$M_0$ and $M_X$ encode the Landau-Zener tunneling  probabilities at each avoided crossing and have the concrete forms
\begin{equation}\label{M0matrix}
    M_0 = \begin{pmatrix}
1 & 0 & & & \\
0 & 1 - P_{12} & 1 - P_{12} & & \\
 & P_{12} & 1-P_{12} & \ddots & \\
 & & \ddots & \ddots & P_{N-1,N} \\
 & & & P_{N-1, N} & 1 - P_{N-1, N}
\end{pmatrix},
\end{equation}
\begin{equation}\label{MXmatrix}
    M_X = \begin{pmatrix}
1-P_{01} & P_{01} & & & \\
P_{01} & 1 - P_{01} & 0 & & \\
 & 0 & 1-P_{23} & \ddots & \\
 & & \ddots & \ddots & 0 \\
 & & & 0 & 1
\end{pmatrix}.
\end{equation}
Each row of $M_0$ and $M_X$ is normalized so that probability is conserved at the crossing: an exciton arriving at a gap either tunnels to the adjacent band with probability $P_{mn}$ or continues in the same band with probability $1-P_{mn}$. Eqs. \eqref{SolutionPaper}, \eqref{M0equationPaper} and \eqref{MXequationPaper} provide unique solutions for $\vec{f}^{(\pm)}(K_x)$ throughout the Brillouin zone, which in turn can be substituted into Eq. \ref{SigmaFinalExp} to compute $\sigma_{xy}^{(CF)}$.
\end{widetext}

Further details regarding the computation of the counterflow conductivity matrix may be found in Appendix \ref{AppendixBoltzmann}.
At this point we turn to our numerical results.

\begin{figure}[t]
    \centering
    \text{(a)}\\
    \includegraphics[width=0.48\textwidth]{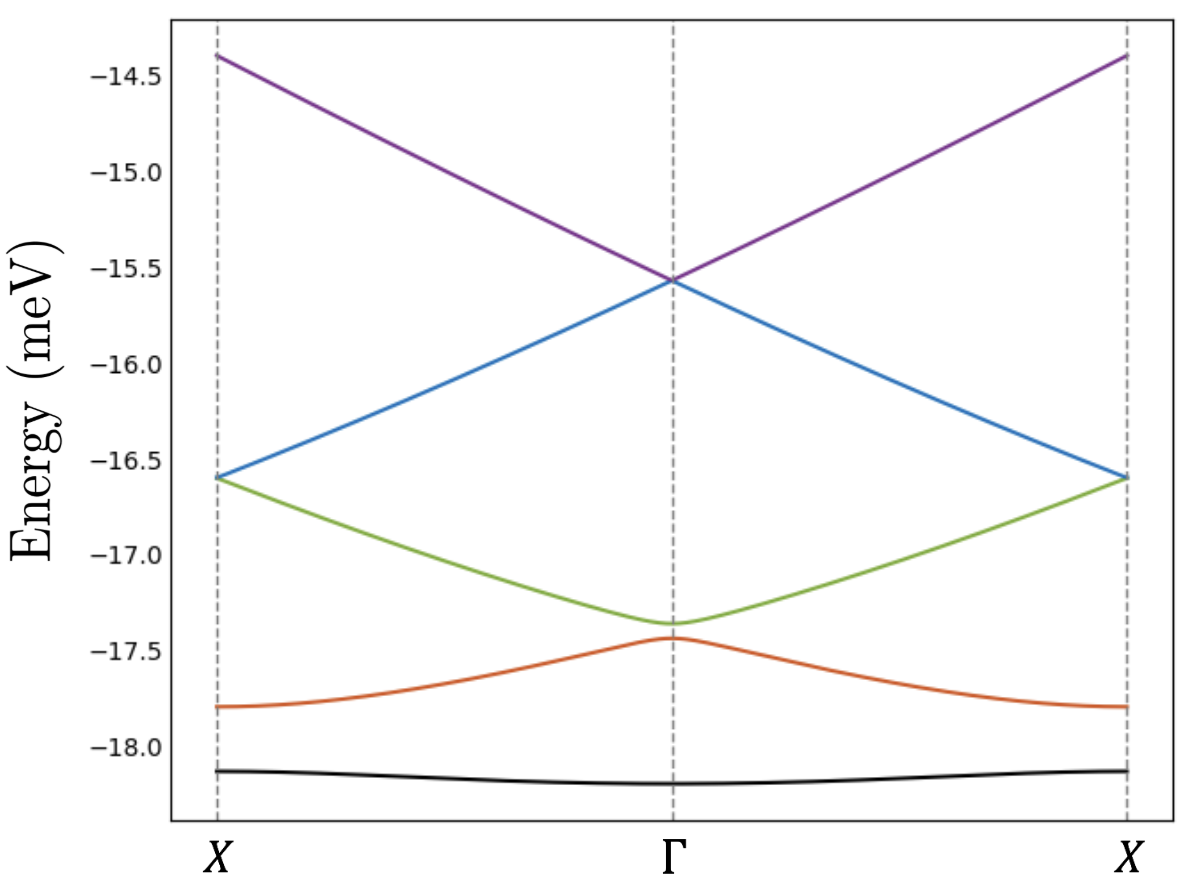}
    \vspace{0.5em}

    \text{(b)}\\
    \includegraphics[width=0.48\textwidth]{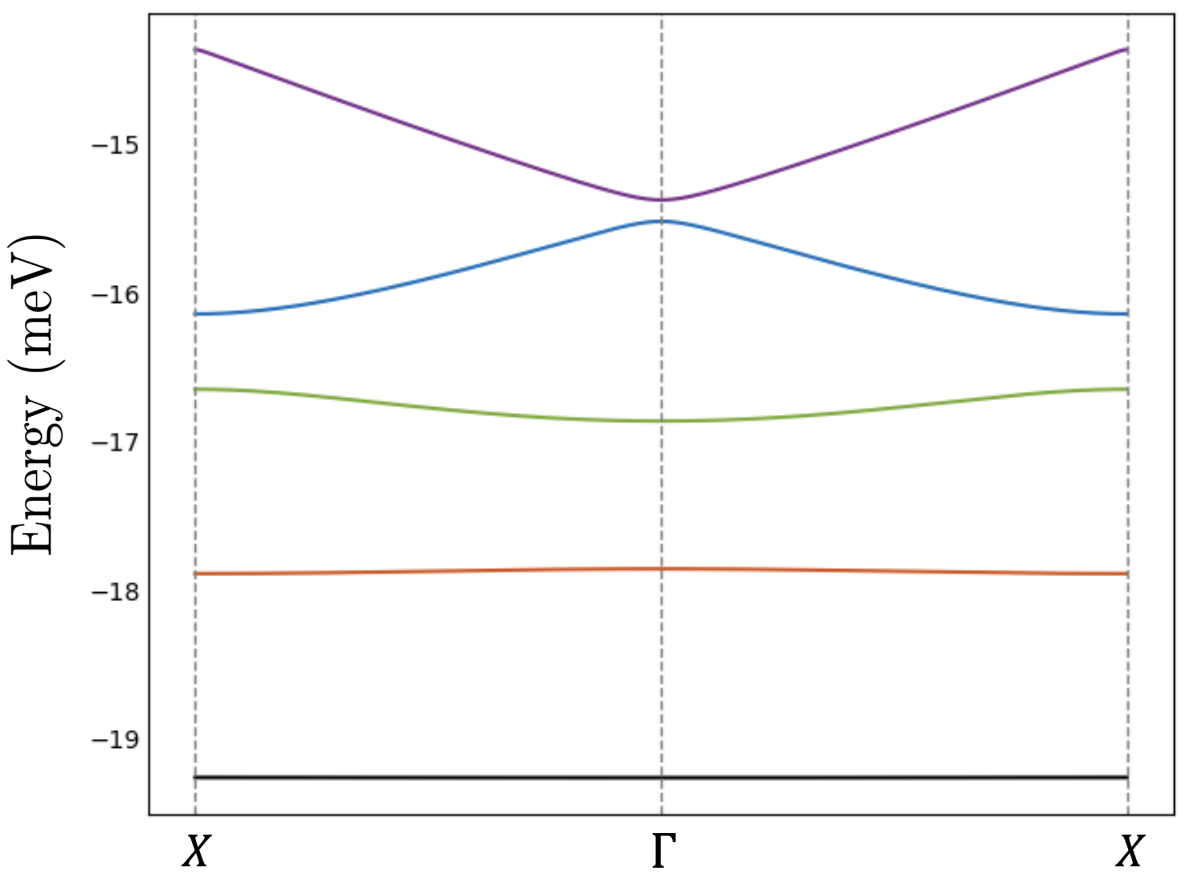}
    \caption{Band structure $E_{n}(K_x)$ for $K_y = 0$ for the lowest five bands of the magnetoexciton for (a) small periodic potential strength $W = 0.09$ meV and (b) large periodic potential strength $W = 0.5$ meV. The dashed vertical lines correspond to the avoided crossings at $K_x = 0$ and $K_x = \pm \pi/a$. As $W$ increases, the gaps open significantly and the lowest band becomes dispersionless. The parameters are: $\ell = 100 \si{\angstrom}$, $d = 3\,\text{nm}$, $a = 140\,\text{nm}$, $\kappa = 3.76$ and $N = 19$.}
    \label{fig:BandStructureW}
\end{figure}

\begin{figure}[t]
    \centering
    \text{(a)}\\
    \includegraphics[width=0.48\textwidth]{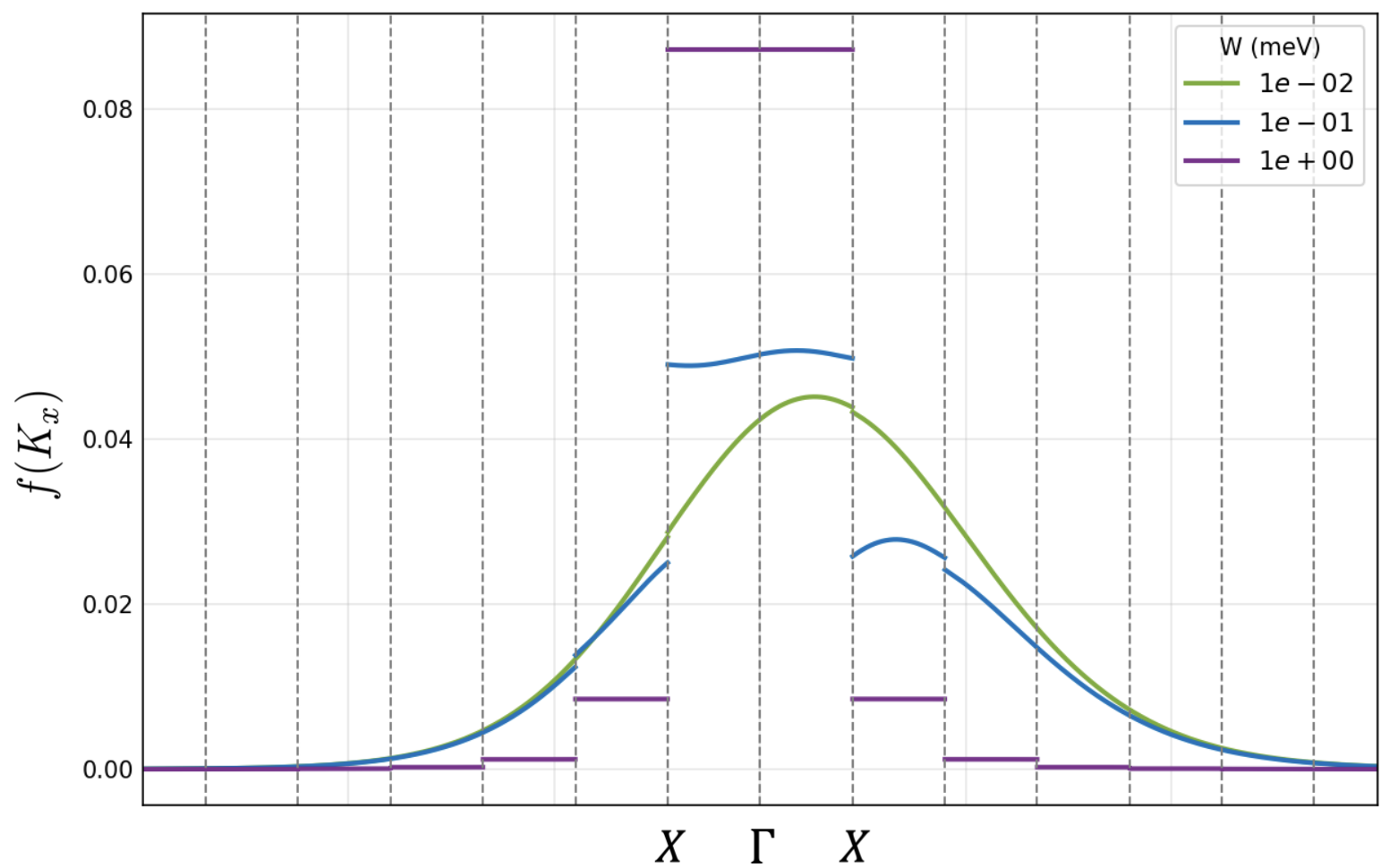}
    \vspace{0.5em}

    \text{(b)}\\
    \includegraphics[width=0.48\textwidth]{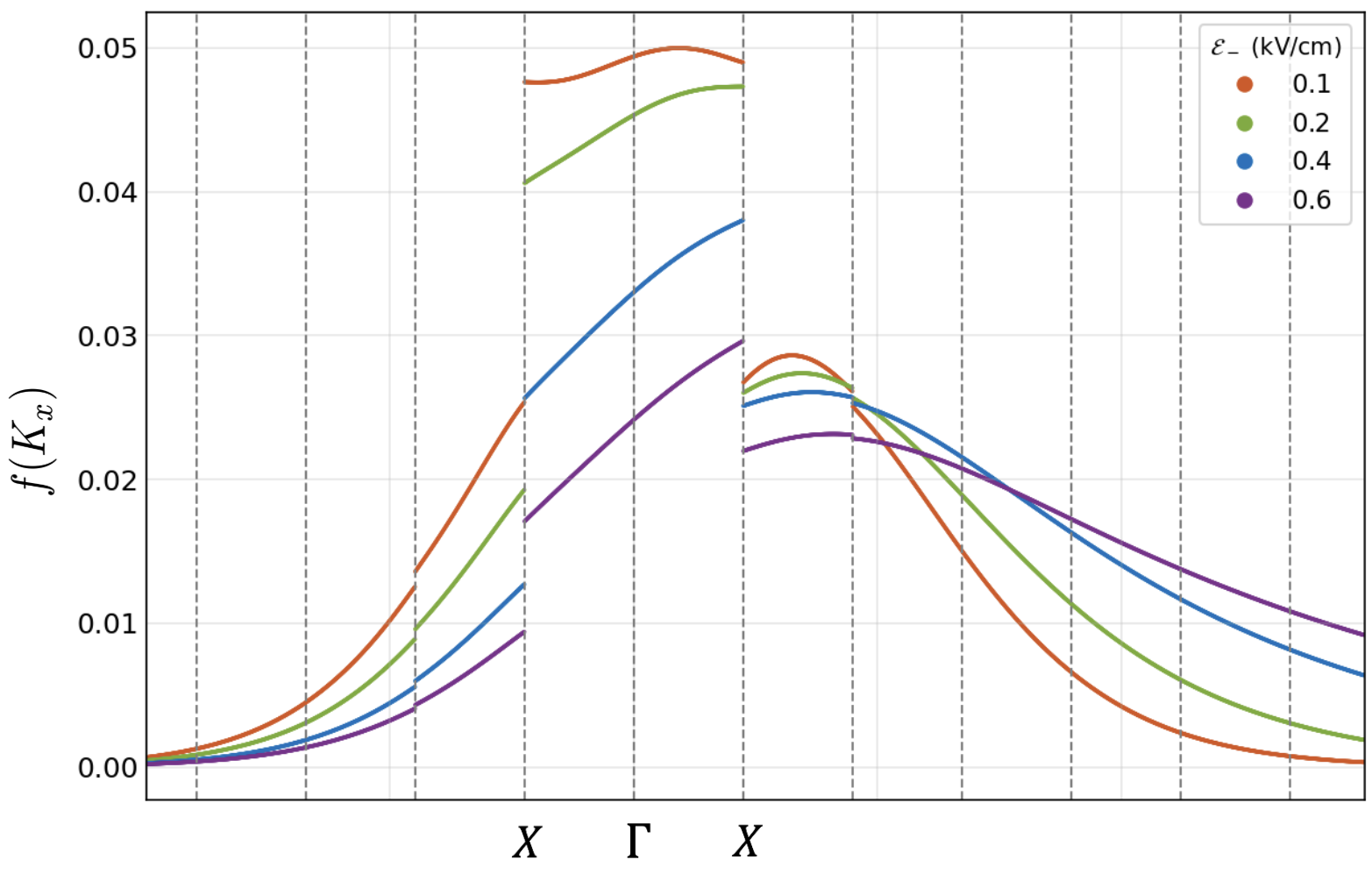}
    \caption{Boltzmann distribution function in the extended zone scheme, for (a) $\mathcal{E}_{-} = 0.1 \,\text{kV/cm}$ while varying different values of the periodic potential strength $W$, and (b) $W = 0.09$ meV for different values of the antisymmetric electric field $\mathcal{E}_{-}$. Other parameters are $T = 10\,\text{K}$, $\ell = 100\,\si{\angstrom}$, $d = 3\,\text{nm}$, $\kappa =3.76$, $a = 140\,\text{nm}$, $\tau = 1\,\text{ps}$, $N = 19$ and $n_{exc} = 10^{9}\,\text{cm}^{-2}$. Note discontinuities at avoided crossings, due to interband tunneling.}
    \label{fig:boltzmannfunction}
\end{figure}

\section{Numerical Results}
\label{ResultsandDisc}


{We begin by examining the exciton band structure $E_{n}(K_x)$ at $K_y =0$, shown in Fig. \ref{fig:BandStructureW} for two values of the periodic potential strength $W$. The electronic structure provides the foundation for all subsequent results: the avoided crossing gaps that govern the Landau-Zener tunneling, the band dispersion that enters the steady state Boltzmann transport equation, and the band-projected QGD that enters Eq. \ref{SigmaFinalExp}.  We assume that the systems we consider are in strong enough magnetic fields and that the carriers are of low enough density that we can assume they reside in the lowest Landau level.  For $W = 0$ the exciton energy takes the form \cite{fertig1989energy}}
\begin{equation}\label{ExcitonEnergyW0}
     E^{(0)}({\bf{K}}) = -\frac{e^2}{\kappa}\int_{0}^{\infty}dq \,e^{-\frac{1}{2}q^2\ell^2}e^{-qd}J_{0}(qK\ell^2),
\end{equation}
where $J_0$ is a Bessel function of the first kind, $e$ is the electron charge, $d$ is the interlayer separation, $\ell$ is the magnetic length, and $\kappa$ is the dielectric constant.

Typical results for $W>0$ are illustrated in Fig.~\ref{fig:BandStructureW}.
For small $W$, the band structure can be understood largely in terms of zone folding. In the absence of the periodic potential, the magnetoexciton dispersion increases from its minimum at ${\bf K} =0$ and asymptotically approaches a constant value at large $|{\bf K}|$, reflecting the vanishing of the electron-hole binding energy as the pair separation grows. The introduction of a unidirectional periodic potential with reciprocal lattice vector ${\bf{g}} = 2\pi\hat{x}/a$ folds this dispersion into the first Brillouin zone $K_x \in [-\pi/a,\pi/a]$, producing multiple bands indexed by the orbital integer $m$ with unperturbed energies $E^{(0)}(K_x + mg_x,K_y)$. At the zone center and the zone boundaries, for fixed $K_y$ bands from adjacent Brillouin zones become degenerate. The periodic potential lifts these degeneracies, opening avoided crossings whose magnitudes grow with $W$. For small $W$, the resulting gaps are small and the bands nearly cross, as is apparent in Fig.~\ref{fig:BandStructureW} (a), particularly at high energy.  As $W$ increases, the gaps become well-resolved at higher energies; moreover, the low-energy bands become increasingly flat.  An example is illustrated in Fig.~\ref{fig:BandStructureW} (b).

With this information, we can compute the nonequilibrium Boltzmann distribution function and the resulting counterflow conductivity. 
Typical results for $\{f_n(K_x)\}$ are presented in Fig. \ref{fig:boltzmannfunction} in the extended zone scheme for fixed $K_y=0$, for different potential strengths $W$ [panel (a)] and for different values of $\mathcal{E}_{-}$ [panel (b)] for a system at temperature $T=10$ K and magnetic field $B=10$ T.  As is evident in Fig. \ref{fig:boltzmannfunction} (a), for $W\lesssim 0.01 \,$ meV the distribution functions are nearly identical to that of $W=0$, indicating that the periodic potential has little effect on the exciton distribution below this potential scale. As $W$ increases above this value, visible jumps open at the avoided crossings located at $K_x=0$ $(\Gamma)$ and $K_x=\pm \pi/a$ $(X)$, and the exciton population increasingly favors the lower over the higher bands, signaling a crossover towards a single-band regime, in which higher bands contribute little to the exciton flow.  Moreover, for $W = 1$ meV and above, the distribution within each band becomes essentially independent of $K_x$ (for fixed $K_y$), so that even highly occupied bands contribute little to the overall counterflow conductivity. This leads to significant suppression of $\sigma_{xy}^{(CF)}$ (see below).

The flattening of the distribution function as $W$ increases has a direct physical origin in the band structure itself. The periodic potential in Eq. \eqref{potential} opens increasingly large gaps at the avoided crossings and simultaneously flattens the dispersions of the lowest energy bands (see Fig. \ref{fig:BandStructureW}). The large gaps effectively confine the excitons to single bands, and {their lack of energetic dispersion effectively makes all momenta equivalent from the perspective of the equations of motion, resulting in the nearly uniform occupation within a band.} 

By contrast, raising $\mathcal{E}_{-}$ from small values tends to reduce the effects of gaps in the spectrum.  This is illustrated in Fig. \ref{fig:boltzmannfunction} (b), where one sees the jumps in the distribution tending to close as   $\mathcal{E}_{-}$ increases. Since the total exciton number density is conserved, the lowest band becomes progressively depleted as excitons are driven into higher bands via Landau-Zener tunneling.
Here the underlying physics is the increasing probability of interband tunneling as the slew rate increases. In addition, the distribution overall becomes increasingly asymmetric, in a manner reminiscent of the non-equilibrium Boltzmann distribution for electrons in a strong driving field \cite{stanton_nonequilibrium_1987}. 


\begin{figure}[t]
    \centering
    \includegraphics[width=0.48\textwidth]{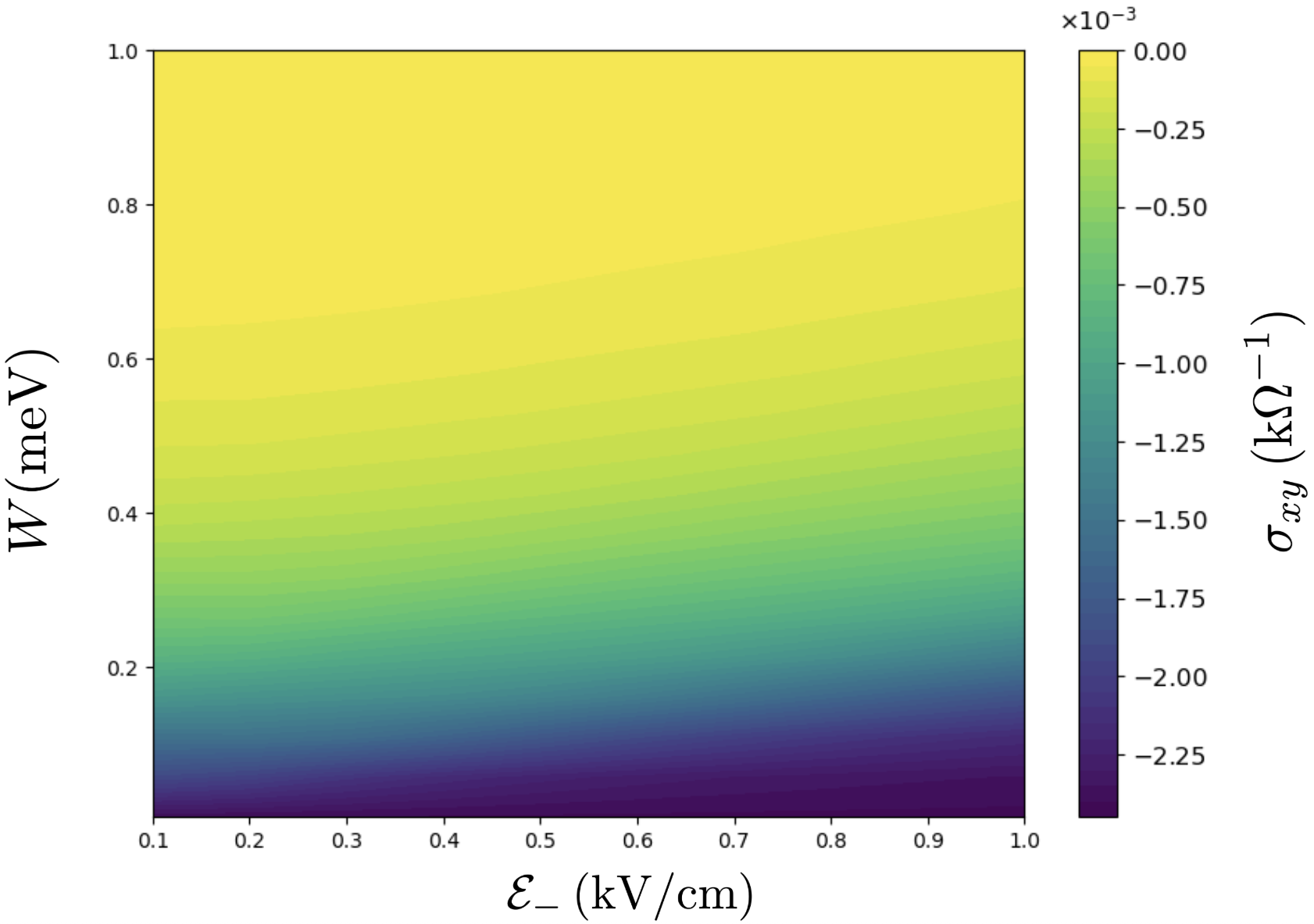}
    \vspace{0.5em}
    \includegraphics[width=0.48\textwidth]{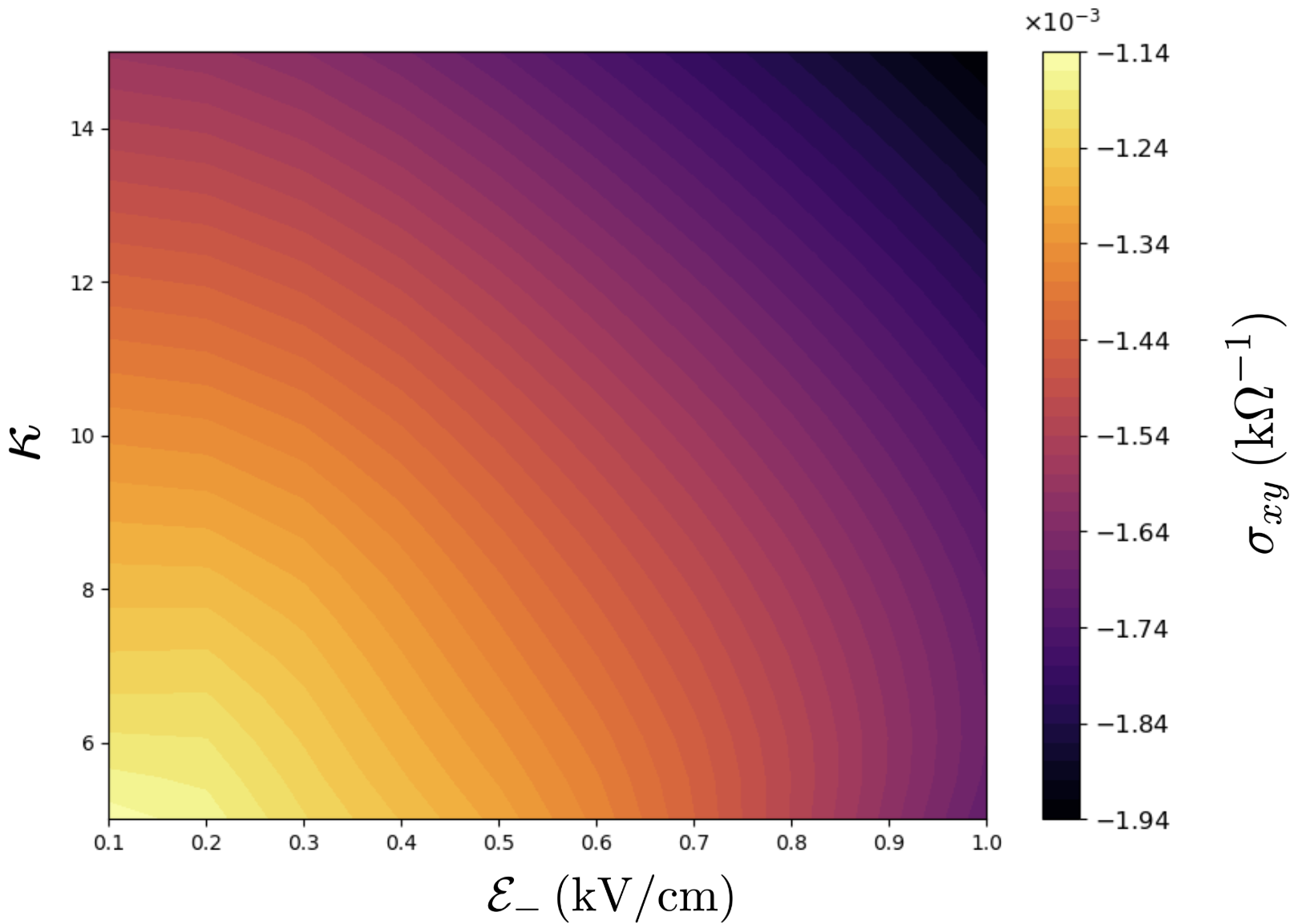}
    \caption{Counterflow conductivity $\sigma_{xy}^{(CF)}$ $(\text{k}\Omega^{-1})$ as a function of (a) the periodic potential strength $W$ and the antisymmetric electric field $\mathcal{E}_{-}$ at fixed dielectric screening constant $\kappa = 3.76$, and (b) $\kappa$ and $\mathcal{E}_{-}$ at fixed $W = 0.17$ meV. For both (a) and (b), $N = 40$, $n_{exc} = 10^{10}\,\text{cm}^{-2}$, and the remaining parameters are the same as Fig. \ref{fig:boltzmannfunction}.}
    \label{fig:Condsigmaxy}
\end{figure}

Fig.~\ref{fig:Condsigmaxy} illustrates typical results for $\sigma^{(CF)}_{xy}$ that result from our numerical computations of the QGD and the non-equilibrium Boltzmann distribution.
Panel (a) shows results from Eq.~\eqref{SigmaFinalExp} as a function of both $W$ and $\mathcal{E}_{-}$ for $T = 10\,\text{K}$, $\ell = 100\,\si{\angstrom}$, $d = 3\,\text{nm}$, $\kappa =3.76$, $a = 140\,\text{nm}$, $\tau = 1\,\text{ps}$, $N = 19$ and $n_{exc} = 10^{10}\,\text{cm}^{-2}$. 
Several features are notable.  As $W \rightarrow 0$, $\sigma_{xy}^{(CF)}$ approaches $-\sigma_{yx}^{(CF)}$, recovering the results expected for a uniform  system, as expected. As $W$ increases, the crossover from a dispersive to a flat-band regime and the relatively larger occupation of the lowest band, tends to suppress $\sigma_{xy}^{(CF)}$.  This may be understood as a reflection of the suppressed $\mathcal{D}_{y}$ magnitudes 
that accompanies the energetic flattening of the lowest energy bands. By contrast, increasing $\mathcal{E}_{-}$ while keeping $W$ fixed tends to {\it increase} the magnitude of $\sigma_{xy}^{(CF)}$. 
The increased Landau-Zener tunneling and resulting enhancement of higher band occupation brings in the somewhat greater slopes in the QGD evident in Fig.~\ref{fig:QGD}(a) as one moves to higher energies, which tends to increase the magnitude of $\sigma_{xy}^{(CF)}$ (see Eq. \eqref{SigmaFinalExp}.) 

Fig.~\ref{fig:Condsigmaxy} (b) shows $\sigma_{xy}^{(CF)}$ as a function of $\kappa$ and $\mathcal{E}_{-}$ at fixed $W = 0.17$ meV. Increasing the dielectric constant $\kappa$ decreases the electron-hole interaction relative to other energy scales in the problem, effectively weakening the effect of their correlated positions. Several features are notable. ({\it i}) Increasing $\kappa$ has an effect analogous to increasing $W$: it flattens the QGD at small momentum (see Fig.~\ref{fig:QGDkappa} in Appendix \ref{AppendixBandStructure}); however, ({\it ii}) as $\kappa$ increases, more bands become occupied since the reduced electron-hole binding energy lowers the energy cost of populating higher bands. For the parameters associated with Fig. \ref{fig:Condsigmaxy} (b), this effect is larger than ({\it i}) above, leading to a net increase in the magnitude of the counterflow conductivity.  ({\it iii}) As $\mathcal{E}_{-}$ increases at fixed $\kappa$, the Landau-Zener tunneling drives the excitons into higher bands and the magnitude of $\sigma_{xy}^{(CF)}$ grows,  as was the case in Fig.~\ref{fig:Condsigmaxy} (a). 


\section{Summary and Discussion}
\label{SecConclusion}






Excitons generically may be characterized by a quantum geometric dipole, an internal polarization that is tied to the structure of the exciton Hilbert space.
In this work we have developed a theory of counterflow conductivity $\sigma_{xy}^{(CF)}$ in a system of interlayer excitons, with the goal of identifying features that provide information on this QGD.  As a basic example, we have focused on a system of interlayer magnetoexcitons in a strong magnetic field and in a unidirectional periodic potential, inducing a simple exciton band structure.  To model transport
we developed a semiclassical theory using a Boltzmann distribution function that takes into account both the multiband structure and the Landau-Zener tunneling that may occur between bands in a driven system. Our modeling demonstrates that broad features of the QGD may be inferred from the behavior of $\sigma_{xy}^{(CF)}$, in particular a tendency to be suppressed in the lowest-energy bands when the periodic potential is sufficiently large, and a tendency to be enhanced at relatively high energies.  In this way, counterflow conductivity offers a window on the internal quantum geometric structure of excitons. 

Not all features of the QGD are captured in a clear way by $\sigma_{xy}^{(CF)}$.  Most prominent are the sharp spikes as a function of $K_x$ at the avoided crossings (see Fig.~\ref{fig:QGD}).  For most bands, these features occupy a very small region of momentum space relative to the full Brillouin zone, in practice yielding only a small contribution to  Eq.~\eqref{SigmaFinalExp}.  Better probes of these features could involve co-flow electrical conductivity, which directly probes relative motion of the electrons and holes.  A particularly interesting quantity in this context could be frequency-dependent noise \cite{stanton_nonequilibrium_1987}, which would capture the swift turnaround of the QGD as it passes through an avoided crossing as a high-frequency component.  Related to this, one might consider whether high-frequency radiation could be detected as excitons move through the band structure and their dipole moments oscillate.  The sharp changes in QGD would be expected to lead to high harmonics of the Bloch oscillation frequency \cite{fahimniya_synchronizing_2021} associated with motion through the band structure.

Our model does not include the possibility of generation and recombination of excitons, as well any inhomogeneity in the exciton density in real space.  For this reason, the analysis applies most directly to systems in which electron and hole carriers are injected by contacts in the separate layers, and relax into exciton states through inelastic scattering processes.  Generalizations of what we have done could incorporate regions of inhomogeneous exciton density, as well as the impact of effectively unbound carriers near contact regions.  This would involve generalizing our Boltzmann approach to model non-trivial real-space distributions.   Incorporation of generation and recombination would also allow more direct modeling of geometries in which the excitons are generated by laser excitation.

The platform we have considered includes only (magneto-)excitons as current carriers in the two layers.  The most directly applicable platforms for such situations would be double layers in which one layer is doped into an electron-like Landau level, and the other into a hole-like Landau level.  In principle, graphene and TMD materials could offer such environments.  By contrast, one can consider quantum Hall bilayers of total filling factor 1, with a bias such that one layer hosts a nearly filled Landau level with a small density of missing electrons, while the other layer hosts the depleted electrons. With such strong biases, the system is not expected to Bose condense \cite{eisenstein_macdonald_2004}; however, it is quite interesting to consider how Bose condensation would affect the behavior of the system.  In addition, a fuller analysis of the $\nu=1$ bilayer quantum Hall system would require incorporation of the ``exciton vacuum'' -- a filled Landau level -- into the computation of $\sigma_{xy}^{(CF)}.$

Finally, our choice of excitons in a strong magnetic field as a platform for QGD effects in exciton transport was made because of the relative simplicity it allows in computing the exciton bands and their associated QGD's.  Other platforms are also relevant, particularly excitons in moir\'e lattice systems which can host relatively large QGD's \cite{yang2026giant}.  In addition to offering a rich and tunable environment for excitons, the nature of these lattices will likely host behaviors stemming from their two-dimensionality which are not present in our simple unidirectional lattice system.  Moreover, it will be interesting to examine whether topological indices may be present in the exciton band structures that impact transport in either the one- or two-dimensional platforms.  We leave these questions for future work.

\section{Acknowledgements}
 H.A.F. and F.I.M. would like to thank Kaijie Yang and P. Myles Eugenio for comments and discussions. Computational work was supported in part by the Lilly Endowment, Inc., through its support for the Indiana University Pervasive Technology Institute. This work was supported by the NSF through Grant No. DMR-2531425, and by the IU Research Emerging Frontiers program. L.B. acknowledges financial support from the Ministerio de Ciencia e Innovaci\'{o}n through the grant PID22024-161156NB-I00. L.B. also acknowledges Severo Ochoa Centres of Excellence program through Grant CEX2024-001445-S.

\onecolumngrid
\appendix

\section{Exciton Band Structure and Multiband QGD}
\label{AppendixBandStructure}
In this Appendix, we provide details on the construction of the exciton band structure and the band-projected QGD in the presence of the one-dimensional 
periodic potential defined in Eq.~\eqref{potential}. In the absence of the periodic potential, exciton states are labeled by the total exciton momentum ${\bf{K}}$ and take the form
\begin{equation}\label{excitonWF}
    \ket{\Phi_{{\bf{K}}}} = \sum_{{\bf{q}}}C_{{\bf{q}}}({\bf{K}})e^{iK_{x}q_{y}\ell^2}\phi_{n_{h},{\bf{q}}}^{(+)}({\bf{r}}_1)\phi_{n_{e},{\bf{K-q}}}^{(-)}({\bf{r}}_2),
\end{equation}
where $\phi^{(+)}_{n_h,\mathbf{q}}(\mathbf{r}_1)$ and $\phi^{(-)}_{n_e,\mathbf{K}-\mathbf{q}}(\mathbf{r}_2)$ 
are the single-particle wavefunctions provided in Ref. \cite{cao_quantum_2021} corresponding to the hole and electron residing in Landau levels 
$n_h$ and $n_e$ respectively, $\ell = \sqrt{\hbar/eB}$ is the magnetic length, and $C_{{\bf{q}}}(\mathbf{K})$ are expansion coefficients, determined by the two-body eigenvalue equation. We work in the strong field limit, where both the electron and hole reside in the lowest Landau level (LLL).

The unperturbed magnetoexciton energy in the LLL is given by 
\cite{fertig1989energy}
\begin{equation}
    E^{(0)}({\bf{K}}) = -\frac{e^2}{\kappa}\int_{0}^{\infty} dq \,e^{-\frac{1}{2}q^2\ell^2}e^{-qd}J_{0}(qK\ell^2),
\end{equation}
where $J_0$ is a Bessel function of the first kind, $e$ is the electron charge, $d$ is the interlayer separation, and $\kappa$ is the dielectric constant. 
This expression reflects the Coulomb binding energy of the electron-hole pair in the lowest Landau level, which decreases monotonically in magnitude as $K$ increases and asymptotically approaches zero at large $K$, reflecting the vanishing of the 
binding energy as the pair separation grows~\cite{kallin_excitations_1984}.

In the presence of the one-dimensional periodic potential
\begin{equation}\label{potentialAppendix}
     V({\bf{r}}_1,{\bf{r}}_2) = W\sum_{i=1}^{2}\cos({\bf{g}}\cdot{\bf{r}}_{i}),
\end{equation}
where  ${\bf{g}} = 2\pi\hat{x}/a$,
the exciton Hamiltonian acquires off-diagonal 
matrix elements that mix states at momenta ${\bf{K}}$ and 
${\bf{K}} \pm {\bf{g}}$.  The matrix 
elements of the periodic potential between unperturbed exciton 
states are
\begin{equation}\label{MatrixElement}
    \braket{\Phi_{{\bf{K}}}|V|
    \Phi_{{\bf{K}}+{\bf{g}}}} = W e^{-(g_x\ell)^2/4}
    \left(1 + e^{-ig_x K_y\ell^2}\right) \equiv \beta(K_y).
\end{equation}
The Hamiltonian takes a 
tridiagonal form in the basis of momentum states. We seek the exciton band structure by constructing the 
variational wavefunction for the $n^{\text{th}}$ band as an 
expansion over unperturbed states shifted by reciprocal 
lattice vectors:
\begin{equation}\label{VariationalWF}
    \ket{\Psi_{n,{\bf{K}}}} = \sum_{m=-N_c}^{N_c}
    c_m^{(n)}({\bf{K}})\ket{\Phi_{{\bf{K}}+m{\bf{g}}}},
\end{equation}
where $N_c$ is a cutoff band retained in the 
expansion and the coefficients $c_m^{(n)}({\bf{K}})$ are 
determined variationally by minimizing the energy 
$\braket{\Psi_{n,{\bf{K}}}|H|\Psi_{n,{\bf{K}}}}$. This leads 
to the eigenvalue equation
\begin{equation}\label{EigenEq}
    \sum_{m'=-N_c}^{N_c} H_{mm'}({\bf{K}})
    c_{m'}^{(n)}({\bf{K}}) = E_n({\bf{K}})c_m^{(n)}({\bf{K}}),
\end{equation}
where the Hamiltonian matrix elements are
\begin{equation}\label{HamiltonianMatrix}
    H_{mm'}({\bf{K}}) = \begin{cases}
         E^{(0)}({\bf{K}}+m{\bf{g}}) & \text{if } m = m', \\
        \beta(K_y) & \text{if } m' = m+1, \\
        \beta^*(K_y) & \text{if } m' = m-1, \\
        0 & \text{otherwise.}
    \end{cases}
\end{equation}
The matrix 
$H_{mm'}({\bf{K}})$ is thus an $N\times N$ tridiagonal 
Hermitian matrix with $N = 2N_c + 1$, which we diagonalize 
numerically to obtain the band energies $E_n({\bf{K}})$ and 
the expansion coefficients $c_m^{(n)}({\bf{K}})$ for each 
band $n$ and momentum ${\bf{K}}$ in the first Brillouin zone.

Our interest is in the evolution of the quantum geometric dipole (QGD) 
in the presence of a periodic potential as the exciton traverses the 
Brillouin zone. In this Appendix, we follow closely 
Ref.~\cite{cao_quantum_2021} in order to derive the band-projected 
QGD in the presence of a unidirectional periodic potential for the general multiband 
case. We define the band-projected cell-periodic functions constructed from Eq. \eqref{VariationalWF}
\begin{equation}\label{GenCellPer}
    \ket{u_{n,{\bf{K}}},\alpha} \equiv 
    e^{-i\left[\alpha{\bf{r}}_1+(1-\alpha){\bf{r}}_2\right]\cdot{\bf{K}}}
    \ket{\Psi_{n,{\bf{K}}}},
\end{equation}
where $\alpha = 1$ and $\alpha = 0$ label the hole and electron 
constituents of the exciton, respectively. These allow us to define 
the band-projected Berry connections for each constituent in accordance with 
Ref.~\cite{cao_quantum_2021}:
\begin{equation*}
    \boldsymbol{\mathcal{A}}_{n}^{(h)}({\bf{K}}) \equiv 
    i\braket{u_{n,{\bf{K}}},1|\nabla_{{\bf{K}}}|u_{n,{\bf{K}}},1},
\end{equation*}
\begin{equation}
    \boldsymbol{\mathcal{A}}_{n}^{(e)}({\bf{K}}) \equiv 
    i\braket{u_{n,{\bf{K}}},0|\nabla_{{\bf{K}}}|u_{n,{\bf{K}}},0},
\end{equation}
with the band-projected QGD formally defined as their difference
\begin{equation}
    \boldsymbol{\mathcal{D}}_n({\bf{K}}) \equiv 
    \boldsymbol{\mathcal{A}}_n^{(h)}({\bf{K}}) - 
    \boldsymbol{\mathcal{A}}_n^{(e)}({\bf{K}}).
\end{equation}
To compute this, we introduce the generalized overlap matrix element
\begin{equation}\label{GenOverlapEq}
    \widetilde{\Gamma}_{\alpha,n}({\bf{K}}_1,{\bf{K}}_2) \equiv 
    \braket{u_{n,{\bf{K}}_1},\alpha|u_{n,{\bf{K}}_2},\alpha},
\end{equation}
which allows us to express the difference of the Berry's connections as 
\begin{equation}\label{ADiff}
    \boldsymbol{\mathcal{A}}_n^{(h)}({\bf{K}}_1) - 
    \boldsymbol{\mathcal{A}}_n^{(e)}({\bf{K}}_1) = 
    i\lim_{{\bf{K}}_2\rightarrow{\bf{K}}_1}
    \nabla_{{\bf{K}}_2}\left[\widetilde{\Gamma}_{1,n}({\bf{K}}_1,{\bf{K}}_2) 
    - \widetilde{\Gamma}_{0,n}({\bf{K}}_1,{\bf{K}}_2)\right].
\end{equation}
Substituting Eq.~\eqref{GenCellPer} into Eq.~\eqref{GenOverlapEq}, we find
\begin{equation}\label{GenOverlap0}
    \widetilde{\Gamma}_{0,n}({\bf{K}}_1,{\bf{K}}_2) = 
    \sum_{m=-N_c}^{N_c} c_m^{*(n)}({\bf{K}}_1)c_m^{(n)}({\bf{K}}_2)\,
    \Gamma_{0}({\bf{K}}_1+m{\bf{g}},{\bf{K}}_2+m{\bf{g}}),
\end{equation}
\begin{equation}\label{GenOverlap1}
    \widetilde{\Gamma}_{1,n}({\bf{K}}_1,{\bf{K}}_2) = 
    \sum_{m=-N_c}^{N_c} c_m^{*(n)}({\bf{K}}_1)c_m^{(n)}({\bf{K}}_2)\,
    \Gamma_{1}({\bf{K}}_1+m{\bf{g}},{\bf{K}}_2+m{\bf{g}}),
\end{equation}
where $\Gamma_0$ and $\Gamma_1$ are the overlaps analogous to Eq. \ref{GenOverlapEq} for the uniform system (see 
Ref.~\cite{cao_quantum_2021}), and cross terms between different values 
of $m$ vanish because states at momenta ${\bf{K}}+m{\bf{g}}$ and 
${\bf{K}}+m'{\bf{g}}$ with $m \neq m'$ are orthogonal when $W=0$. Taking the 
gradient of the difference of the overlaps with respect to ${\bf{K}}_2$,
and substituting into Eq.~\eqref{ADiff}, we obtain
\begin{equation}\label{GenQGDFinal}
    \boldsymbol{\mathcal{A}}_n^{(h)}({\bf{K}}_1) - 
    \boldsymbol{\mathcal{A}}_n^{(e)}({\bf{K}}_1) = 
    i\sum_{m=-N_c}^{N_c}|c_m^{(n)}({\bf{K}})|^2\,
    \lim_{{\bf{K}}_2\rightarrow{\bf{K}}_1}\nabla_{{\bf{K}}_2}
    \Bigl[\Gamma_1({\bf{K}}_1+m{\bf{g}},{\bf{K}}_2+m{\bf{g}}) 
    - \Gamma_0({\bf{K}}_1+m{\bf{g}},{\bf{K}}_2+m{\bf{g}})\Bigr].
\end{equation}
Recognizing that the quantity inside the sum is precisely the QGD of 
a uniform magnetoexciton evaluated at momentum 
${\bf{K}}+m{\bf{g}}$~\cite{cao_quantum_2021}, we arrive at the 
central result of this Appendix,
\begin{equation}\label{MultiQGDFinalGen}
    \boldsymbol{\mathcal{D}}_n({\bf{K}}) = 
    \sum_{m=-N_c}^{N_c}|c_m^{(n)}({\bf{K}})|^2\,
    \boldsymbol{\mathcal{D}}_{ME}({\bf{K}}+m{\bf{g}}),
\end{equation}
where $\boldsymbol{\mathcal{D}}_{ME}({\bf{q}}) \equiv 
{\bf{q}}\times\hat{z}\,\ell^2$ is the QGD of a 
magnetoexciton for a uniform system~\cite{cao_quantum_2021, fertig_many-body_2025}. 

\begin{figure}[h]
    \centering
    \includegraphics[width=0.48\textwidth]{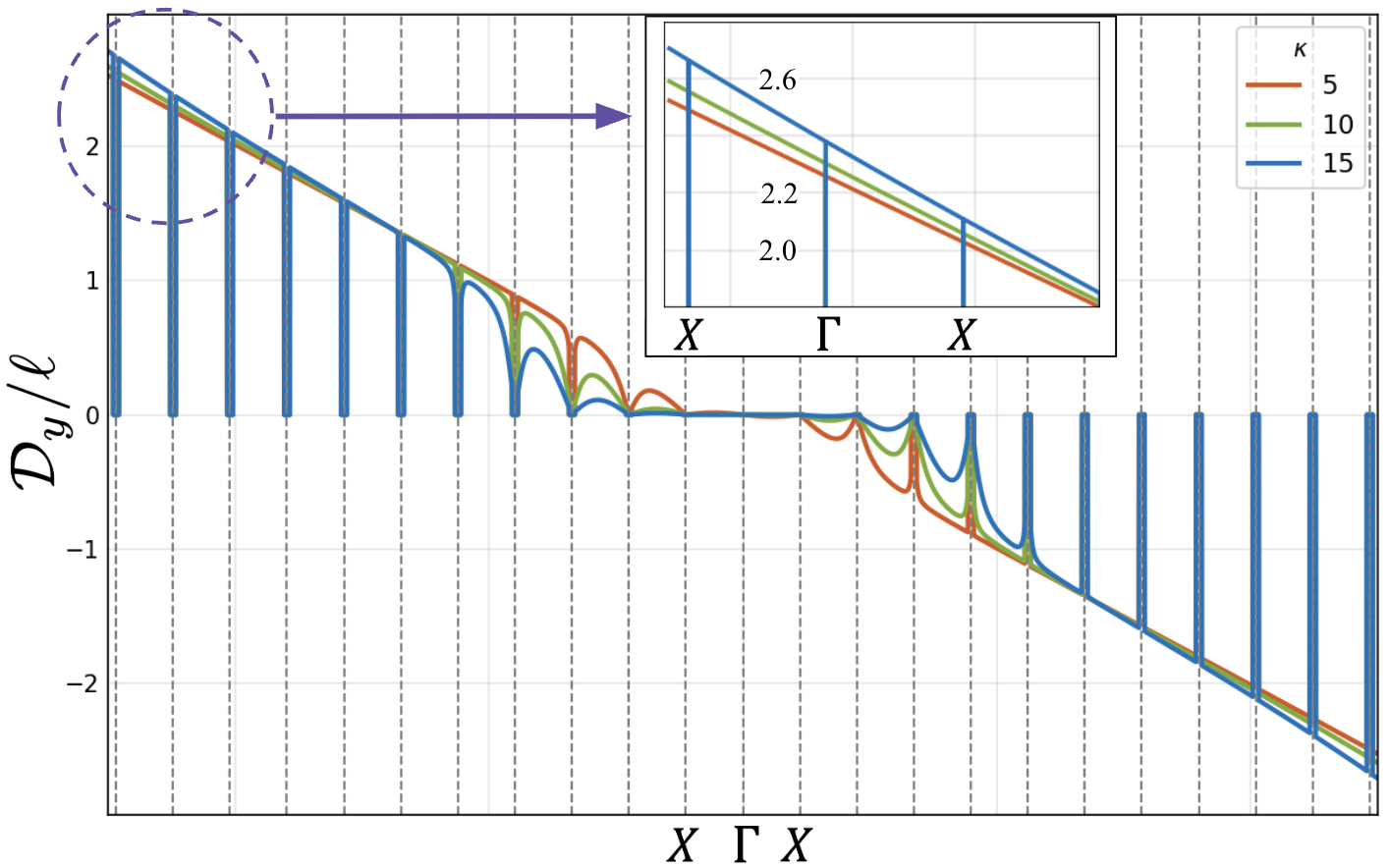}
    \caption{Band-projected QGD for fixed $K_y = 0$ for several values of the dielectric constant $\kappa$ in the extended zone scheme. The dashed vertical lines indicate avoided crossings where the QGD vanishes. Inset: Behavior of the QGD for large momentum magnitude in extended zone scheme. Parameters used were $\ell = 100 \,\si{\angstrom}$, $d = 3 \, \text{nm}$, $a = 140 \, \text{nm}$ , $W = 0.17$ meV and $N = 41$.}
    \label{fig:QGDkappa}
\end{figure}

Results for the QGD at different values of $W$ are presented in the main text.  To conclude this Appendix, we illustrate the effect of the dielectric constant $\kappa$ on the band-projected QGD. Fig.~\ref{fig:QGDkappa} shows $\mathcal{D}_{y}(K_x)$ at fixed $K_y = 0$ for several values of $\kappa$. As $\kappa$ increases, the Coulomb interaction between the 
electron and hole is increasingly screened, which modifies the unperturbed magnetoexciton dispersion $ E^{(0)}({\bf{K}})$ and consequently the expansion coefficients $c_m^{(n)}({\bf{K}})$. The result is a 
suppression and flattening of the band-projected QGD at small momentum, reflecting the reduced electron-hole binding and the redistribution of spectral weight among the shifted momentum states ${\bf{K}}+m{\bf{g}}$. 
This behavior is analogous to increasing $W$ in that both suppress the QGD gradient $\partial\mathcal{D}_{n,y}/\partial K_x$ that enters the counterflow conductivity Eq.~\eqref{SigmaFinalExp}. 
Competing with this,
for high bands the QGD grows in magnitude with increasing $\kappa$ faster than would be expected based purely on zone-folding, in a way analogous to what happens with increasing $W$ at fixed $\kappa$: the contribution from states of higher momenta to the exciton wavefunctions tends to increasingly outweigh those of lower momenta as the band index (and exciton energy) increases, giving extra weight to contributions with larger QGD.

\section{Counterflow Conductivity}
\label{AppendixBoltzmann}


In this Appendix, we provide details of the derivation of the 
counterflow conductivity tensor $\overleftrightarrow{\sigma}^{(CF)}$ 
starting from the semiclassical equations of motion. This is constructed from the 
exciton number current, which induces electric currents in opposite directions in the two layers, leading to an electric counterflow current
\begin{equation}\label{cfcurrent}
    {\bf{j}}_{+}^{(CF)} = e\sum_{n}\int_{BZ} 
    \frac{d^2K}{(2\pi)^2}f_{n}({\bf{K}})\dot{{\bf{R}}}_{+,n},
\end{equation}
where $\dot{{\bf{R}}}_{+,n}$ is the velocity of an exciton 
center-of-mass coordinate ${\bf{R}}_+$ in band $n$, whose evolution is given by 
the equations of motion Eq.~\eqref{EOMposition}, and 
$f_n({\bf{K}})$ is the nonequilibrium distribution function 
obtained from the Boltzmann equation described in 
the main text. Since we are interested in the contribution of the QGD to 
transport, the form of Eq.~\eqref{EOMposition}
suggest we should consider the \textit{deviation} of the current from its 
value in the absence of the symmetric field $\boldsymbol{\mathcal{E}}_+$,
\begin{equation}\label{DevCFcurrent}
    \delta {\bf{j}}_{+}^{(CF)} \equiv 
    {\bf{j}}_{+}^{(CF)}(\boldsymbol{\mathcal{E}}_{+}) - 
    {\bf{j}}_{+}^{(CF)}(\boldsymbol{\mathcal{E}}_{+} = 0) = 
    \frac{e^2}{\hbar}\sum_{n}\int_{BZ} 
    \frac{d^2K}{(2\pi)^2}f_{n}({\bf{K}})
    \nabla_{{\bf{K}}}[\boldsymbol{\mathcal{E}}_{+}
    \cdot \boldsymbol{\mathcal{D}}_{n}({\bf{K}})].
\end{equation}
Note that the 
distribution function $f_n({\bf{K}})$ depends on 
$\boldsymbol{\mathcal{E}}_-$ but not on 
$\boldsymbol{\mathcal{E}}_+$, since the antisymmetric 
field drives the exciton momentum while the symmetric 
field couples only to the QGD. This separation is what 
makes $\delta{\bf{j}}_+^{(CF)}$ a linear response to 
$\boldsymbol{\mathcal{E}}_+$.

Decomposing Eq.~\eqref{DevCFcurrent} into component 
form, valid whether 
$\boldsymbol{\mathcal{E}}_+$ is static or dynamic,
\begin{equation}
    \delta j_{+,\nu}^{(CF)} = \sum_{\mu}
    \mathcal{E}_{+,\mu}(t)\Biggl\{\frac{e^2}{\hbar}
    \sum_{n}\int_{BZ} \frac{d^2K}{(2\pi)^2}
    f_{n}({\bf{K}})\frac{\partial\mathcal{D}_{n,\mu}
    ({\bf{K}})}{\partial K_\nu}\Biggr\}.
\end{equation}
Identifying the quantity in braces with the conductivity 
tensor via $\delta j_{+,\nu}^{(CF)} = \sum_\mu 
\sigma_{\nu\mu}^{(CF)}\mathcal{E}_{+,\mu}$, we arrive 
at the general expression for the counterflow conductivity,
\begin{equation}\label{CFconduct}
    \sigma_{\nu\mu}^{(CF)} = \frac{e^2}{\hbar}
    \sum_{n}\int_{BZ}\frac{d^2K}{(2\pi)^2}
    f_{n}({\bf{K}})\frac{\partial \mathcal{D}_{n,\mu}}
    {\partial K_{\nu}}, \quad \nu,\mu = x,y,
\end{equation}
as displayed in the main text. 
It shows that the counterflow conductivity is determined 
by two ingredients: the nonequilibrium distribution 
function $f_n({\bf{K}})$, which encodes the response 
of the exciton population to $\boldsymbol{\mathcal{E}}_-$, 
and the gradient of the band-projected QGD 
$\partial\mathcal{D}_{n,\mu}/\partial K_\nu$, which 
encodes the quantum geometric structure of the exciton 
band. 

In what follows, we evaluate each component of 
$\overleftrightarrow{\sigma}^{(CF)}$ using the explicit 
form of the band-projected QGD derived in 
Appendix~\ref{AppendixBandStructure}.
Recalling that $\boldsymbol{\mathcal{D}}_{ME}({\bf{q}}) = {\bf{q}}\times\hat{z}\,\ell^2$, the two components of the 
band-projected QGD are
\begin{equation}\label{DxComponent}
    \mathcal{D}_{n,x}({\bf{K}}) = 
    \sum_{m=-N_c}^{N_c}|c_m^{(n)}({\bf{K}})|^2\,K_y\ell^2 
    = K_y\ell^2,
\end{equation}
and
\begin{equation}\label{DyComponent}
    \mathcal{D}_{n,y}({\bf{K}}) = 
    -\sum_{m=-N_c}^{N_c}|c_m^{(n)}({\bf{K}})|^2\,
    (K_x + mg_x)\ell^2 
    = -\ell^2\left(K_x + g_x\sum_{m=-N_c}^{N_c}
    m\,|c_m^{(n)}({\bf{K}})|^2\right),
\end{equation}
where we used the normalization condition 
$\sum_{m}|c_m^{(n)}|^2 = 1$. Note that the $x$-component 
$\mathcal{D}_{n,x} = K_y\ell^2$ is independent of 
both the band index $n$ and $K_x$, while the 
$y$-component retains a nontrivial $n$ and 
$K_x$ dependence through the weighted sum 
$\sum_m m|c_m^{(n)}({\bf{K}})|^2$.

From Eq.~\eqref{DxComponent}, 
$\mathcal{D}_{n,x}({\bf{K}}) = K_y\ell^2$ is 
independent of $K_x$ and $n$, and its derivative with 
respect to $K_y$ is simply
\begin{equation}
    \frac{\partial \mathcal{D}_{n,x}}{\partial K_y} 
    = \ell^2.
\end{equation}
Substituting into Eq.~\eqref{CFconduct}, one obtains
\begin{equation}
    \sigma_{yx}^{(CF)} = \frac{e^2\ell^2}{\hbar}
    \sum_n\int_{BZ}\frac{d^2K}{(2\pi)^2}f_n({\bf{K}}) 
    = \frac{e^2\ell^2}{\hbar}n_{exc},
\end{equation}
where we used $\sum_n\int_{BZ}\frac{d^2K}{(2\pi)^2}
f_n({\bf{K}}) = n_{exc}$, the total exciton number 
density \cite{girvin2019modern}. This recovers Eq.~\eqref{SigmaYX} of the 
main text, and holds for any distribution function 
$f_n({\bf{K}})$, independently of the periodic 
potential strength $W$ or the driving field 
$\mathcal{E}_-$.

For $\sigma_{xy}^{(CF)}$, we require 
$\frac{\partial\mathcal{D}_{n,y}}{\partial K_x}$. 
Differentiating Eq.~\eqref{DyComponent} with respect 
to $K_x$ yields
\begin{equation}
    \frac{\partial\mathcal{D}_{n,y}}{\partial K_x} = 
    -\ell^2\left(1 + g_x\sum_{m=-N_c}^{N_c}
    m\frac{\partial|c_m^{(n)}({\bf{K}})|^2}
    {\partial K_x}\right),
\end{equation}
which depends nontrivially on ${\bf{K}}$ through the 
$K_x$-dependence of the expansion coefficients 
$c_m^{(n)}({\bf{K}})$. These are determined by the 
diagonalization of the tridiagonal Hamiltonian matrix 
described in Appendix~\ref{AppendixBandStructure} and 
must be evaluated numerically. Substituting into 
Eq.~\eqref{CFconduct} yields
\begin{equation}
    \sigma_{xy}^{(CF)} = \frac{e^2}{\hbar}\sum_{n}
    \int_{BZ}\frac{d^2K}{(2\pi)^2}f_{n}({\bf{K}})
    \frac{\partial \mathcal{D}_{n,y}}{\partial K_{x}}.
\end{equation}
In our computations we evaluate this numerically using the band-projected 
QGD (Appendix~\ref{AppendixBandStructure}) and the nonequilibrium distribution function 
$f_n({\bf{K}})$ obtained from the Boltzmann equation 
as described in in the main text.

We now demonstrate that $\sigma_{xx}^{(CF)}$ vanishes 
exactly. From Eq.~\eqref{CFconduct} we have
\begin{equation}
    \sigma_{xx}^{(CF)} = \frac{e^2}{\hbar}\sum_n
    \int_{BZ}\frac{d^2K}{(2\pi)^2}f_n({\bf{K}})
    \frac{\partial\mathcal{D}_{n,x}}{\partial K_x}.
\end{equation}
Eq.~\eqref{DxComponent},tells us that
$\mathcal{D}_{n,x}({\bf{K}}) = K_y\ell^2$ is 
independent of $K_x$ for all bands $n$. Therefore,
\begin{equation}
    \frac{\partial\mathcal{D}_{n,x}}{\partial K_x} 
    = 0,
\end{equation}
from which it immediately follows that
$
    \sigma_{xx}^{(CF)} = 0.
$


Finally we consider 
$\sigma_{yy}^{(CF)}$, which requires 
$\frac{\partial\mathcal{D}_{n,y}}{\partial K_y}$.  In the absence of the periodic potential, this quantity vanishes identically because the QGD in this situation is purely transverse.  In the presence of the periodic potential,
differentiating Eq.~\eqref{DyComponent} with respect 
to $K_y$ yields
\begin{equation}
    \frac{\partial\mathcal{D}_{n,y}}{\partial K_y} = 
    -\ell^2 g_x\sum_{m=-N_c}^{N_c}
    m\frac{\partial|c_m^{(n)}({\bf{K}})|^2}
    {\partial K_y}.
\end{equation}
There is a weak $K_y$-dependence in the 
eigenvector coefficients $|c_m^{(n)}({\bf{K}})|^2$, 
making $\frac{\partial|c_m^{(n)}|^2}{\partial K_y}$ 
small but generically nonzero. 
For example, typical values of $|\sigma_{yy}^{(CF)}/\sigma_{yx}^{(CF)}|$ are less than $10^{-5}$.
For this reason we focus on $\sigma_{xy}^{(CF)}$ and  $\sigma_{yx}^{(CF)}$ in the 
main text.

\twocolumngrid

\bibliographystyle{apsrev4-2}
\bibliography{references}

\end{document}